\documentclass[usenatbib]{mn2e}

\usepackage{graphicx}
\usepackage{multirow}
\usepackage{url}
\usepackage{gensymb}

\newcommand{\mpcoh}{\,h^{-1}\,{\rm Mpc}}

\title[Large-scale redshift-space distortions]
  {Interpreting large-scale redshift-space distortion measurements}

\author[L. Samushia et al.]{L.~Samushia,$^{1,2}$\thanks{email:
lado.samushia@port.ac.uk} W.~J.~Percival,$^1$
and A.~Raccanelli$^1$\\
$^1$ Institute of Cosmology and Gravitation, University of Portsmouth, Dennis
Sciama Building, Portsmouth P01 3FX, U.K.\\
$^2$ National Abastumani Astrophysical Observatory, Ilia State University, 2A
Kazbegi Ave., GE-1060 Tbilisi, Georgia\\}

\begin{document}

\maketitle

\begin{abstract}
  The simplest theory describing large-scale redshift-space
  distortions (RSD), based on linear theory and distant galaxies,
  depends on the growth of cosmological structure, suggesting that
  strong tests of General Relativity can be constructed from galaxy
  surveys. As data sets become larger and the expected constraints
  more precise, the extent to which the RSD follow the simple theory
  needs to be assessed in order that we do not introduce systematic
  errors into the tests by introducing inaccurate simplifying
  assumptions. We study the impact of the sample geometry, non-linear
  processes, and biases induced by our lack of understanding of the
  radial galaxy distribution on RSD measurements. Using LasDamas
  simulations of the Sloan Digital Sky Survey II (SDSS-II) Luminous
  Red Galaxy (LRG) data, these effects are shown to be important at
  the level of 20 per cent. Including them, we can accurately model
  the recovered clustering in these mock catalogues on scales
  $30$--$200\mpcoh$. Applying this analysis to robustly measure
  parameters describing the growth history of the Universe from the
  SDSS-II data, gives $f(z=0.25)\sigma_8(z=0.25)=0.3512\pm0.0583$ and
  $f(z=0.37)\sigma_8(z=0.37)=0.4602\pm0.0378$ when no prior is imposed
  on the growth-rate, and the background geometry is assumed to follow
  a $\Lambda$CDM model with the WMAP + SNIa priors. The standard WMAP
  constrained $\Lambda$CDM model with General Relativity predicts
  $f(z=0.25)\sigma_8(z=0.25)=0.4260\pm0.0141$ and
  $f(z=0.37)\sigma_8(z=0.37)=0.4367\pm0.0136$, which is fully consistent with
  these measurements.
\end{abstract}

\begin{keywords}
  gravity --- cosmological parameters --- dark energy --- large-scale structure of Universe 
\end{keywords}

\section{Introduction}
\label{sec:intro}

The statistical quantification of Redshift-Space Distortions (RSD)
provides a robust method for measuring the growth of structure on very large
scales. RSD arise because we infer galaxy distances from their redshifts using
the Hubble law: the radial component of the peculiar velocity of individual
galaxies will contribute to each redshift and be misinterpreted as being
cosmological in origin, thus altering our estimate of the distances to them. The
measured clustering of galaxies will therefore be anisotropic and the additional
radial signal can be used to determine the characteristic amplitude of the
pair-wise distribution of the peculiar velocities at a given scale, which in
turn depends on the growth rate.

Many previous analyses have used RSD to measure the cosmological
growth rate using both the correlation function and power spectrum
(see, for example,
\citealt{hawkins03,percival04,zehavi05,tegmarketal06,guzzo08,cabre09,song10}).
In general these studies used clustering information over a small
range of scales, and a simplified modelling procedure in order to make
the measurements.

Large-scale RSD measurements provide results that can be compared to
direct measurements of peculiar velocities in the local Universe: both
observations depend on the amplitude of the velocity field. Recent
analyses of the local data seem to indicate the presence of
unexpectedly large bulk flows, 2$\sigma$ higher than $\Lambda$CDM
predictions \citep{watkins09,feldman10,macaulay11}, although
\citet{nusser10} present more compatible measurements. Large bulk
velocities were also detected through measurements of the kinetic
Sunyaev-Zeldovich effect on the X-ray cluster catalog
\citep{kashlinsky09}. The excess velocities detected are at odds with
the previous RSD measurements from the 2-degree Field Galaxy Redshift
Survey (2dFGRS; \citealt{colless03}) and the Sloan Digital Sky Survey
(SDSS; \citealt{york00}) discussed above, which give results broadly
consistent with $\Lambda$CDM models. There is therefore strong
motivation for considering if systematic effects could be affecting
either set of observations.

If we assume that observed galaxies are sufficiently far away that
their separations are small compared to the distances between them and
the observer (the ``plane-parallel'' approximation) then, to linear
order, the relationship between the redshift-space galaxy
power-spectrum $P_{\rm gg}^{\rm s}$, the real-space matter
power-spectrum $P_{\rm mm}^{\rm r}$ and the growth rate is simple
\citep{kaiser87,hamilton97},
\begin{equation}
  \label{eq:rsd} P_{\rm gg}^{\rm s}(k,\mu)=P_{\rm mm}^{\rm r}(k)(b+f\mu^2)^2,
\end{equation} 
where $b$ accounts for a linear deterministic bias between galaxy and
matter overdensity fields, $f$ is the logarithmic derivative of the
growth factor by the scale factor $f\equiv d\ln G/d\ln a$, and $\mu$
is the cosine of the angle to the line-of-sight. In our paper we study
possible theoretical systematics beyond the model of
Eq.~(\ref{eq:rsd}), that could effect the measurements of clustering
on large scales including the effects due to wide-angle corrections,
large-scale nonlinearities, sample geometry and the effects of the
radial model for the distribution of galaxies.

Nonlinear effects change the real-space matter power spectrum, the
velocity power spectrum, the matter--velocity cross-correlation, and
introduce further $\mu$ dependent terms into this expression
\citep{scoccimarro04}. On small scales the dominant nonlinear
contribution comes from the Fingers-of-God (FOG) effect
\citep{jackson72}. FOG arise because within dark matter halos the
velocities of galaxies quickly become virialized and their
power-spectrum is highly nonlinear. This effect can be approximated by
including a phenomenological term in Eq.~(\ref{eq:rsd}) that reduces
power on small scales \citep{PeaDod96} or using a more complicated
expression based on higher order computations in perturbation theory
\citep[see e.g.,][]{scoccimarro04,taruya10}. The phenomenological
damping terms used to describe the FOG effects are not accurate
\citep{scoccimarro04,jennings11} and the results of perturbation
theory are not easy to implement in a computationally fast and
efficient way. The effects of nonlinear growth on the real-space
power-spectrum are also important and difficult to model for an
arbitrary cosmological model. Although in principle these nonlinear
effects can be estimated analytically using perturbation theory,
comparison of different perturbation theory methods to the results of
high-resolution N-body simulations shows that at low redshifts the
range of scales where the perturbation theory is reliable is rather
small \citep{carlson09}. In addition, the assumption that the bias
between matter overdensities and galaxies is linear is not accurate
even for the scales as large as 30$\mpcoh$ \citep{okamura11}.

Wide-angle corrections are needed because, if the angle $\alpha$ that
a galaxy pair forms with respect to the observer is large, the
distance between galaxies is comparable to their distance to the observer and
the ``plane-parallel'' approximation (and hence Eq.~\ref{eq:rsd}) breaks down.
The redshift-space correlation function and the power-spectrum in this case will
also depend on the third variable that could be chosen to be the angle $\alpha$.
The wide-angle linear redshift-space correlation function and power-spectrum as
a function of all three variables have been computed
\citep{zaroubi93,szalay98,szapudi04,matsubara04,papai08}. In fact, the
wide-angle correlation function does not deviate significantly from its
``plane-parallel'' counterpart if the opening angle $\alpha$ is less than
$10^\circ$. In previous work we validated this work by analysing mock galaxy
catalogs \citep{raccanelli10}.
	    
For surveys that cover a significant fraction of the sky, the distribution of
galaxies pairs becomes non-trivial. The survey geometry results in the galaxy
pair distribution that has a complicated dependence on the variables $r$, $\mu$
and $\alpha$, since not all sets of their combinations are equally likely or
even geometrically possible. In particular the distribution of $\mu$ does not
correspond to that of an isotropic pair distribution. This will strongly bias
the measurement of angular momenta of the correlation function, and in fact,
often dominates over differences between ``plane-parallel'' and ``wide-angle''
effects for galaxy pairs with the same $\mu$ \citep{raccanelli10}.

RSD data on very large scales, although in principle available in
current data sets, do not contribute significantly to current data
analyses. The reason for this is twofold: the signal-to-noise of
currently available clustering data becomes low at scales larger than
100$\mpcoh$ so most of the available cosmological information is on
smaller scales; also the large scale clustering measurements are
vulnerable to different observational (improper modelling of seeing,
galactic extinction, etc.) and theoretical systematic effects which,
if not taken into account properly, could strongly bias results of
data analysis. In particular, our ability to model the radial galaxy
distribution accurately can cause strong effects on these
large-scales, and is worthy of further investigation
\citep{percival10,kazin10}. Being able to model these data has many
advantages. First, if accurate measurements are available, more data
will result in stronger constraints on cosmological parameters. In
addition, measurements on large scales are significantly less affected
by the systematics introduced by nonlinear phenomena. Some important
physical processes can be measured only on very large scales. For
example, non-Gaussian initial conditions, if present, will affect the
real-space galaxy clustering on large scales
\citep{dalal08,desjacquesseljak10}, and could be compared against the
RSD signal, which depends on the matter field.

We investigate the significance of these effects by performing an analysis on a
large suite of N-body simulations, testing for systematic effects that could
result in real data giving a signal different from the plane-parallel linear RSD
formula. Using mock samples, we are able to accurately fit the expected
correlation function on scales between $30$--$200\mpcoh$, to a level well below
the statistical error on the measurement from any one sample.

We apply the knowledge learned in this analysis to robustly measure RSD in the
Sloan Digital Sky Survey (SDSS) Data Release 7 (DR7) sample of Luminous Red
Galaxies (LRGs) and measure cosmological parameters describing amplitude and
growth of the perturbations in different models. We find that the accuracy of
SDSS DR7 data is at the threshold where the inclusion of RSD information on
scales larger than 100$\mpcoh$ affects the measurements but does not improve the
result significantly. In addition, we find that the non-linear and survey
geometry effects are significant for this sample.  The next generation of
spectroscopic surveys, such as Baryon Oscillation Spectroscopic Survey ({\it
BOSS}) \citep{schlegel09}, {\it BigBOSS} \citep{schlegeletal09} and {\it Euclid}
\citep{laureijs09} will enable us to measure the clustering of galaxies at large
scales with extremely high accuracy and using RSD data on large scales will be
crucial for constraining growth of structure in the Universe and the nature of
gravity in a robust way. On large scales, ignoring effects such as wide-angles
would lead to systematic deviations. However, these effects are known, so can be
easily included. On small scales, non-linear behavior is harder to model, so
there may be unavoidable additional errors here. 

This paper is organised as follows, in Sec.~\ref{sec:calc} we describe our data
set and the method of measurement. In Sec.~\ref{sec:mocks} we present our
theoretical model for the RSD on very large scales and check its validity with
the results of N-body simulations. In Sec.~\ref{sec:meas} we describe
theoretical models and assumptions made when extracting cosmologically
relevant information. In Sec.~\ref{sec:res} we present results of our data analysis.
In Sec.~\ref{sec:conc} we conclude and discuss the relevance of the contents of
our paper to future spectroscopic galaxy surveys.

\section{Calculating momenta of the correlation function} 
\label{sec:calc}

\subsection{The SDSS data} \label{ssec:sloan}

\begin{figure}
  \includegraphics[height=80mm,width=80mm]{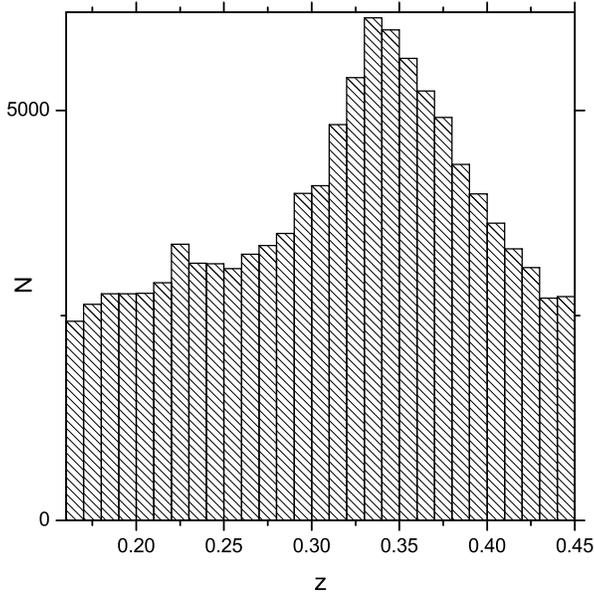}
  \caption{Histogram showing the redshift distribution of galaxies in the SDSS
    DR7 LRG catalog used in our analysis.}
  \label{fig:gdist}
\end{figure}

We use data from the SDSS data release 7 (DR7), which obtained wide-field CCD photometry
(\citealt{C}) in five passbands ($u,g,r,i,z$; e.g., \citealt{F}),
amassing nearly 10,000 square degrees of imaging data for which object
detection is reliable to $r \sim 22$ (\citealt{DR7}).  From these
photometric data, Luminous Red Galaxies (LRG) were targeted
\citep{eisenstein01} and spectroscopically observed, yielding a sample
of 106,341 LRGs in the redshift bin $0.16<z<0.44$. The redshift
distribution of galaxies in this catalog is shown on
Fig.~\ref{fig:gdist}.

To study clustering properties of LRGs we create a random catalog that
has unclustered ``galaxies'' randomly distributed with the same
angular mask as SDSS DR7. The angular distribution of these galaxies
was chosen as described in \citet{reid09}. The method for, and effect
of estimating the expected radial distribution of the galaxies is
described in Section~\ref{ssec:zdist}. Our random catalog has
approximately 50 times more objects than the real catalog.

\subsection{Methodology}  \label{ssec:corr}
  
We assign each galaxy a weight
\begin{equation}
  \label{eq:weight}
  w = \frac{1}{1+n({\bf r})\overline{P}},
\end{equation}
where $n({\bf r})$ is a local density of galaxies in units of
$\left(\mpcoh\right)^{-3}$ in a neighbourhood of the galaxy of
interest located at a position $\bf r$ and $\overline{P}=10000
\left(\mpcoh\right)^3$. This weighting is optimal for the premise that
galaxies Poisson sample the underlying matter field \citep{FKP}. Recent
work has shown that it may be possible to beat this if we can estimate
the mass associated with each galaxy \citep{seljak09}: we do not
attempt this in our work.

In order to extract information about the evolution of structure
growth, we divide the LRG sample in two redshift bins so that the
weighted number of galaxies
\begin{equation}
  \label{eq:wnumber}
  N_{\rm w} = \displaystyle\sum_{ i} w_{ i},
\end{equation}
is approximately equally split between them. The two redshift bins are
$0.16<z<0.32$ and $0.32<z<0.44$ while the effective redshifts for our two
bins are $z=0.25$ and $z=0.37$. Effective redshift is defined as
\begin{equation}
  \label{eq:zeff}
  z_{\rm eff}=\frac{1}{N_w(N_w-1)}\displaystyle\sum_i^N\displaystyle\sum_{j>i}^N
  w_iw_j(z_i + z_j),
\end{equation}
\noindent
where $N_w$ is given by Eq.~(\ref{eq:wnumber}).

For each pair of objects in our galaxy catalog, random catalog or
cross pairs between galaxy and random catalogs, we compute the
distance between objects $r$, the angle $\alpha$ that the objects make
with respect to us, and the cosine of the angle that the bisector of
the angle between the objects makes with the line connecting them
(assuming a flat geometrical model) $\mu$. Due to statistical isotropy
about the observer, these three variables are sufficient to completely
describe the RSD expected for each pair. We bin $r$ in 65 equal
logarithmic bins from $1\mpcoh$ to $200\mpcoh$, $\mu$ in 200 equal
bins from 0 to 1, $\cos(\alpha)$ in 400 bins from -1 to 1 and count the
number of galaxies in each bin.\footnote{Later in the paper we will show that
for this particular observed geometry wide-angle effects are negligible, which
means that for this particular case the $\alpha$-label could have been drop from
the beginning. We still keep the $\alpha$-label in the rest of this paper for
the completeness of formalism.}

To convert angular and redshift separations of galaxies into physical
separations a fiducial cosmological model is needed. We compute distances in a
spatially-flat $\Lambda$CDM fiducial model. If the real geometry of the Universe
is different from the one described by our fiducial model it will bias the
measurements of clustering through the Alcock-Paczynski effect. We discuss this
issue in Sec.~\ref{ssec:ap}.

\begin{figure}
  \includegraphics[height=210mm,width=80mm]{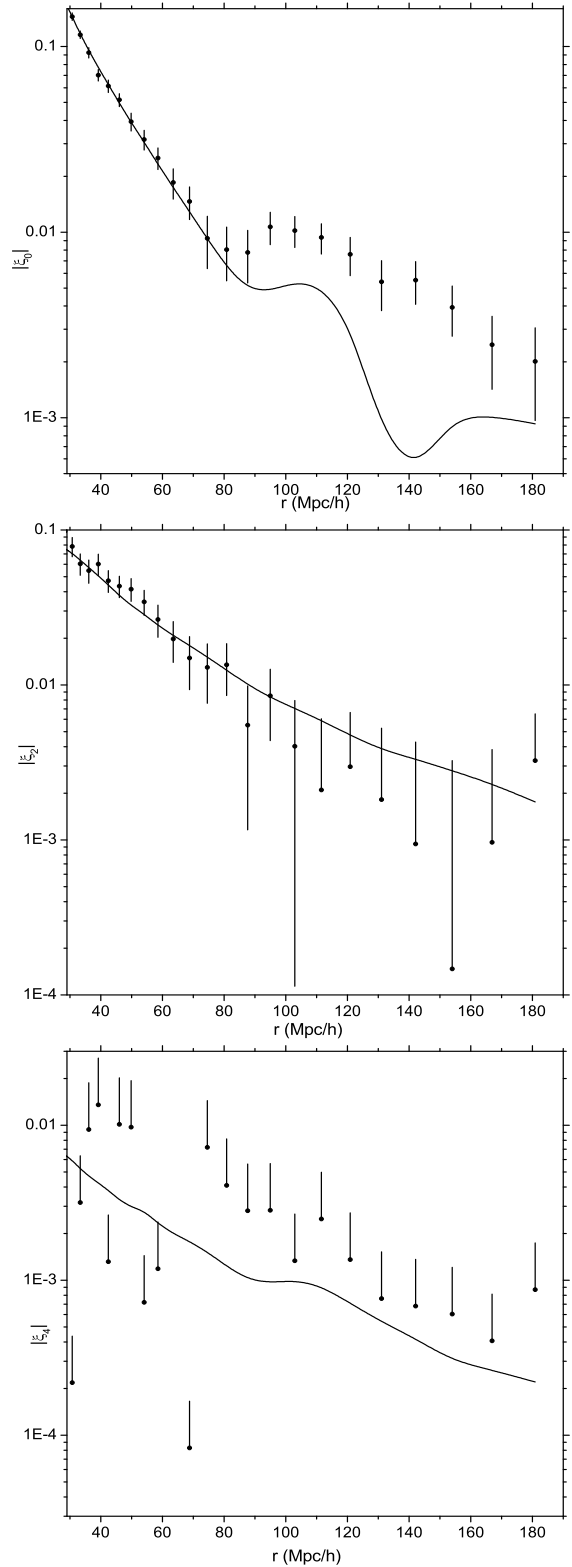}
  \caption{Measurements of $\hat{\xi_\ell}(r)$ from SDSS DR7 LRGs in
    a redshift range $0.16<z<0.44$. The statistical error-bars were calculated as
    described in Section~\ref{sec:mocks} and represent only the diagonal
    elements of the whole covariance matrix. The absence of lower error-bar on
    some measurements indicates that they are consistent with zero. Solid line
    shows a theoretical prediction with the shape corresponding to the best fit
    cosmology to current WMAP and SNIa measurements and the amplitude given by the best-fit
    values to the data.}
  \label{fig:xim}
\end{figure}

In the plane-parallel approximation, all of the available linear RSD
information can be extracted from the zeroth, second and fourth
Legendre momenta of the galaxy correlation function with respect to
the variable $\mu$ \citep{hamilton92}. Given that we expect that
wide-angle and non-linear effects will give relatively small
deviations about this approximation, we should expect that, even in
the more general case, these momenta contain almost all of the
available RSD information. We therefore choose to fit to these
measurements in our work. This will be discussed further in
Section~\ref{sec:mocks}. To estimate those three we use Landy-Szalay
type estimators \citep{landy93}
\begin{eqnarray}
  \hat{\xi_\ell}(r_i) &=& \displaystyle\sum_{j,k}
  \left\{\left[DD(r_i,\mu_j,\alpha_k)-2DR(r_i,\mu_j,\alpha_k)\right.\right. \nonumber\\
  &+&\left.\left. RR(r_i,\mu_j,\alpha_k)\right]P_\ell(\mu_j)\right\}/\displaystyle\sum_{j,k}
RR(r_i,\mu_j,\alpha_k),    
  \label{eq:ls_ell}
\end{eqnarray}
where $DD(r_i,\mu_j,\alpha_k)$, $DR(r_i,\mu_j,\alpha_k)$ and
$RR(r_i,\mu_j,\alpha_k)$ are the numbers of galaxy-galaxy,
galaxy-random and random-random pairs in bins centered on $r_i$,
$\mu_j$ and $\alpha_k$. The $P_\ell$ are the $\ell^{ th}$ Legendre
momenta. 

The measurements of $\hat{\xi_\ell}(r)$ from all LRGs in our catalog, together
with theoretical predictions of spatially-flat $\Lambda$CDM model with
$\Omega_{\rm m}=0.25$ are shown on Fig.~\ref{fig:xim}.  Note that,
Fig.~\ref{fig:xim} shows only statistical errors and does not show extra
systematic errors due to uncertainty in radial distribution of galaxies (for
details on systematic errors see Sec.~\ref{ssec:zdist}). Also the statistical
errors are computed from diagonal elements of the covariance matrix only, the
whole structure of the covariance matrix is such that measurements are more
likely to be systematically above or below theoretical line rather then randomly
scattered around the theoretical prediction as for noncorrelated Gaussian
variables. 

\begin{figure}
  \includegraphics[height=80mm,width=80mm]{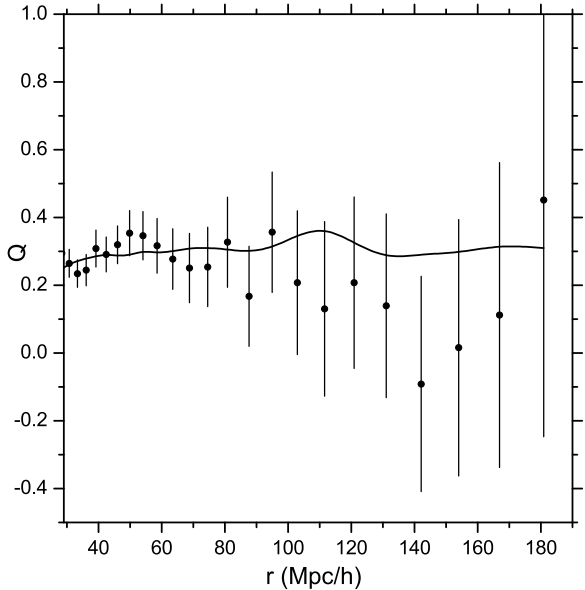}
  \caption{Measurements of $\hat{Q}(r)$ from SDSS DR7 LRGs in a redshift range
  $0.16<z<0.44$. Error-bars were calculated as described in
  Section~\ref{sec:mocks} and show only the diagonal elements of statistical
  covariance matrix. Solid line shows a theoretical prediction with the shape
  corresponding to the best fit cosmology to current WMAP and SNIa measurements
  and the amplitude given by the best-fit values to the data.}
  \label{fig:Qm}
\end{figure}

RSD measurements are also often extracted from the normalised
quadrupole $Q(r)$ \citep{hamilton92}, defined as
\begin{equation}
  \label{eq:Q}
  Q(r) =
  \frac{\xi_2(r)}{\xi_0(r)-\frac{3}{r^3}\displaystyle\int_0^r\xi_0(r')r'^2dr'}.
\end{equation}
We can form an estimator for $Q(r)$ by replacing integral in
Eq.~(\ref{eq:Q}) by a discreet sum
\begin{equation}
  \label{eq:Qm}
  \hat{Q}(r_{ i}) =
  \frac{\hat{\xi_2}(r_{ i})}{\hat{\xi_0}(r_{ i})-\frac{3}{r_{
  i}^3}\displaystyle\sum_{j=0}^{j\le i}\hat{\xi_0}(r_{ j})r_{ j}^2\Delta
    r_{ j}}.
\end{equation}
The measured $Q(r)$ from the SDSS DR7 LRG data is shown in Fig.~\ref{fig:Qm}. 
The details of how the statistical error bars are computed are discussed in
Sec.~\ref{ssec:mocktests}.

\subsection{Modelling the redshift distribution of SDSS LRGs}
\label{ssec:zdist}

 To measure a correlation function from a survey accurately, we must know what
 an unclustered distribution of galaxies would look like in the same volume. The
 unclustered distribution can in principle be derived by averaging observations
 in many unconnected regions. Since the real data covers only a relatively small
 volume the expected galaxy density (in the absence of clustering) is hard to
 determine in this way (cosmic variance). The wrong estimate of unclustered
 distribution will bias the measurements of the correlation function. This
 effect is especially important on large scales where the fluctuations we wish
 to measure are small.

In this paper we compute correlation function by using a spline fit to the
galaxy redshift distribution (with parameters as given in \citealt{percival10}).
We will refer to it as a ``spline'' random catalog. The exact form of the random
catalog will depend on the position and number of nodes used for the spline. In
the limiting case when the number of nodes goes to infinity while the spacing
between nodes goes to zero we will have a random catalog that has exactly the
same redshift distribution of galaxies as data: in effect we assign a randomly
chosen galaxy redshift to the random object . We refer to it as a ``z-shuffled''
catalog. We also construct a random catalog by randomly mixing angular positions
and redshifts in the galaxy catalog (later referred to as ``3D-shuffled''
catalog).  We should expect the shuffled catalogs to remove some structure, as
fluctuations in the galaxy density caused by large-scale structure will be
smoothed. The ``spline'' and ``z-shuffled'' catalogs, unlike the ``3D-shuffled''
catalog, have angular positions of objects choosen at random within the sample
angular mask.

\begin{figure}
  \includegraphics[height=80mm,width=80mm]{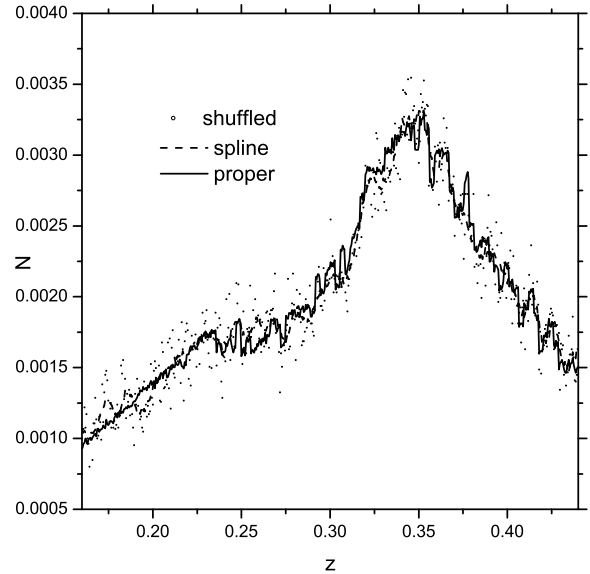}
  \caption{Radial distribution of galaxies in different random catalogs. Dashed
  line corresponds to the ``shuffled'' random catalog, which has radial
  distribution identical to that of data.  Dashed line corresponds to ``spline''
  random catalog and solid line corresponds to ``proper'' random catalog.}   
    \label{fig:zdist2}
\end{figure}

To quantify the possible systematic offset induced by improper modelling of the
radial distribution of galaxies we use large suite of LasDamas N-body
simulations (McBride et al., in prep)  which are designed to replicate the observed
geometry of the SDSS-II (for more details on LasDamas simulations and how we use
them see Sec.~\ref{ssec:mocktests}). For the mock catalogs the unclustered
redshift distribution of galaxies is known and we will refer to the random
catalog based on this known distribution as a ``proper'' random catalog. We
compute correlation function of mocks using the ``proper'', the ``spline'', the
``z-shuffled'' and the ``3D-shuffled'' random catalogs. The radial distribution of
galaxies in each of these random catalogs, for one of the LasDamas mock, is
shown on Fig.~\ref{fig:zdist2}. ``z-shuffled'' and ``3D-shuffled'' catalogs have
identical redshift distribution of points but different angular distribution.

\begin{figure}
  \includegraphics[height=160mm,width=80mm]{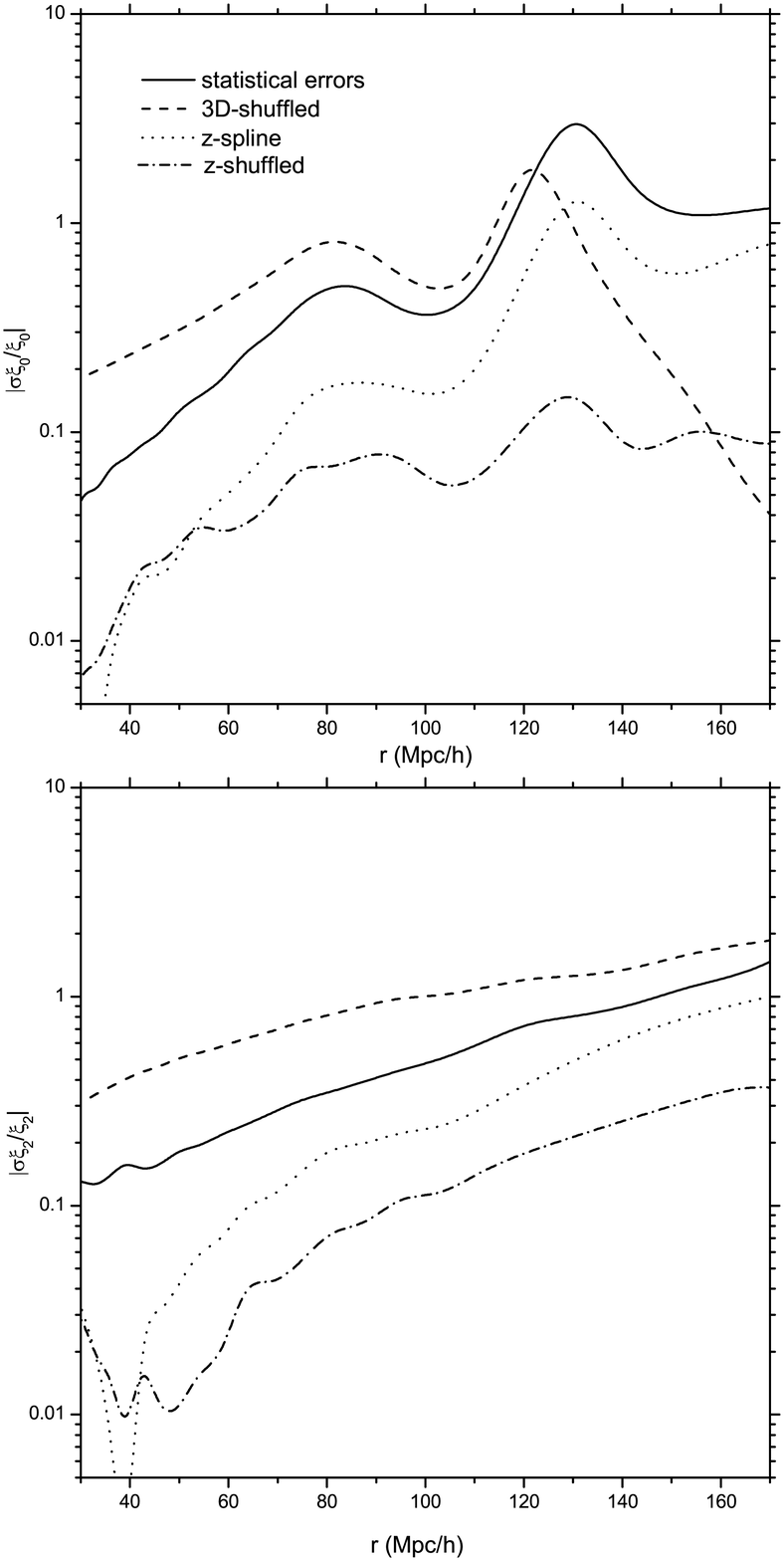}
  \caption{Relative errors introduced in the measurement of the first two even
  Legendre momenta of the correlation function using different random catalog.
  The solid line shows statistical errors.}   
    \label{fig:zdist}
\end{figure}

Figure~\ref{fig:zdist2} shows that although the ``spline'' random catalog
follows the general shape of the ``proper'' random catalog, it does not
reproduce the real radial distribution of galaxies accurately. The induced
systematic errors on measurements of $\xi_0$ and $\xi_2$ when using different
random catalogs catalog with respect to the ``proper'' catalog, averaged over
all 80 LasDamas mock, are shown on Fig.~\ref{fig:zdist} along with the
statistical errors.

Figure~\ref{fig:zdist} shows that when the ``3D-shuffled'' catalog is used the
systematic offset is larger then statistical errors on all scales.  The
statistical errors are larger for $\xi_2$ compared to $\xi_0$.  This is not
surprising since the errors in redshift distribution affect clustering mainly in
radial direction. $\xi_0$ measures average clustering in all direction, while
$\xi_2$ measures the excess of clustering across the line-of-sight compared to
radial and is expected to be affected by this systematics more. For the
``z-shuffled'' and ``splined'' random catalogs the systematic offset is a
fraction then statistical errorbars and ``z-shuffled'' catalog seems to be
performing better compared to the ``splined'' catalog.

In our analysis we use the ``splined'' catalog to compute correlation functions
and ignore this systematic offset since it is small compared to current
errorbars. The exact scaling of the offset with galaxy number density, and
radial fit technique will be investigated in a separate paper.
\subsection{Excess of power on large scales}
\label{sec:excess}

The top panel of Fig.~\ref{fig:xim} shows that there is an excess of power in
the measured monopole of the correlation function with respect to the
predictions of flat WMAP 5 normalised $\Lambda$CDM model. This excess has been
observed previously in spectroscopic \citep{okumura08,sanchez09,cabre09,kazin10}
and photometric \citep{thomas10,crocce11,ross11} data sets. Recent analysis of mock catalogs
resembling SDSS DR7 showed that 6 out of 160 measured correlation functions were
positive up to the scales of 200 $\mpcoh$ \citep[see, Fig.~12 in][]{kazin10}.
This suggests that the excess seen is only mildly statistically significant.

If the signal is physical, the modifications to the standard cosmological model
that result in stronger clustering, such as some modified gravity theories or
clustering dark energy \citep{takada06}, could explain this anomaly. Other
possible explanations are the presence of large non-Gaussian initial conditions
\citep{dalal08} or isocurvature perturbations. It should be noted however that
the amount of non-Gaussianity required to generate such a big offset is ruled
out by other observations \citep[see, e.g.,][]{desjacquesseljakiliev10}.
Possible observational systematics include, for instance, improper modelling of
extinction and seeing. Both of these could introduce spurious extra angular
fluctuations in the data that would later be misinterpreted as an excess of
power. \citet{sanchez09} showed that if there is a systematic constant shift in
measured correlation function (due to calibration errors or evolutionary
effects) this does not bias the estimated best-fit valus of cosmological
parameters significantly.

Inaccuracies in the measured clustering induced by assuming an incorrect
radial distribution, are also large and could in principle explain the anomaly
\citep{kazin10}. The error rescaling suggested in previous section makes the
inconsistency with the standard model on very large scales less
severe.\footnote{Also note that the measurements of spherically averaged
correlation function at different scales are strongly correlated. This makes the
deviations to the one side of the model prediction more probable, the
significance of this deviation being smaller then what it would be for
uncorrelated measurements.}

\section{Modeling RSD on large scales}
\label{sec:mocks}
  
\subsection{Plane-Parallel, Linear model}  \label{ssec:kaiser}

The linear, plane-parallel model for RSD is often termed the Kaiser
model \citep{kaiser87}. In the following, we follow standard convention and
denote the observed galaxy overdensity field by $\delta_{\rm g}$, with a
superscript $s$ in redshift-space and $r$ in real-space. A given Fourier $k$
mode of this overdensity can be expressed to linear order
in overdensity as
\begin{equation}
  \label{eq:stor}
  \delta_{\rm g}^s(k) = \delta_{\rm g}^r(k) - \mu^2\theta_{\rm g}(k),
\end{equation}
where $\mu$ is the cosine of an angle with respect to the line of
sight and $\theta_g = \nabla\cdot {\bf u}$ is a divergence of the galaxy
velocity field. We follow the commonly adopted assumption that this is
equal to the divergence of the matter field, assuming no velocity
bias, i.e. we assume that $b_{\rm v}(k)=1$, where $\theta_{\rm
  g}(k)=b_{\rm v}(k)\theta_{\rm m}(k)$. A subscript ${\rm g}$ shows
that a quantity relates to the galaxy field, and a subscript ${\rm m}$ denotes the
matter field.

The two-point function of this overdensity field is anisotropic,  
\begin{equation}  \label{eq:pgg}
  P_{\rm gg}^s(k) = P_{\rm gg}^r(k) - 2\mu^2P_{\rm g\theta}^r(k) +
    \mu^4P_{\rm \theta\theta}^r(k),
\end{equation}
where $P_{xy}=\langle\delta_x\delta_y\rangle$ denotes a
cross-power-spectrum of fields $x$ and $y$. 

If we further assume that, in real-space, the overdensities in the
galaxy field are linear functions of overdensities in the matter field
$\delta_{\rm g}=b\delta_{\rm m}$ and the velocity divergence can be
related to the matter overdensities using the linearized continuity
equation $\delta_{\theta}=-f\delta_{\rm m}$, then the redshift-space
power-spectrum can be simply expressed in terms of the real-space
power-spectrum as
\begin{equation}  \label{eq:kaiser}
  P_{\rm gg}^s(k,\mu) = (b+f\mu^2)^2P_{\rm mm}^r(k).
\end{equation}
The proportionality constant $b$ between matter and galaxy
overdensities is the bias factor and the coefficient $f$ between
velocity divergence and the matter overdensity is equal to the
logarithmic derivative of the growth factor by the scale factor $d\ln
G/d\ln a$, which follows from the continuity equation combined with
scale-independent growth.

The redshift-space correlation function is given by the Fourier transform of
Eq.~(\ref{eq:pgg})
\begin{equation}
  \label{eq:xigg}
  \xi_{\rm gg}^s(r,\mu) = \displaystyle\int P_{\rm gg}^s(k,\mu)\exp(-i{\bf kr})
  d^3k,
\end{equation}
and can also be expressed in terms of its Legendre momenta
\citep{hamilton92,hamilton97}.

\subsection{Wide-angle effects}  \label{ssec:wa}
  
When the distance between galaxy pairs is comparable to the distance
between galaxies and the observer, the theory of Sec.~\ref{ssec:kaiser} can not be
used to describe RSD effects. The redshift-space correlation
function becomes a function of three variables that can be chosen to
be the separation between galaxies $r$ and the two angles
$\phi_1$ and $\phi_2$ that galaxies form with an arbitrary $z$
axis with respect to the observer
\citep{zaroubi93,szalay98,szapudi04}. \citet{papai08} showed that this
correlation function, when expanded in tripolar spherical harmonics,
\begin{equation}  \label{eq:trip}
  \xi_s(r,\phi_1,\phi_2) =
    \displaystyle\sum_{\ell_1,\ell_2,\ell}B^{\ell_1,\ell_2,\ell}(r,\phi_1,\phi_2)
    S_{\ell_1,\ell_2,\ell}(\hat{\bf x}_1,\hat{\bf x}_2,\hat{\bf x}),
\end{equation}
gives only a few non-zero terms in the absence of an observational
window.  Here $\hat{\bf x}_1$ and $\hat{\bf x}_2$ are the unit vectors in the
direction of two galaxies and $\hat{\bf x}$ is a unit vector pointing in
the direction from galaxy one to galaxy two.

Eq.~(\ref{eq:trip}) can be recast as a function of variables $r$,
$\mu$ and $\alpha$, where $\alpha$ is an angle the galaxies make with
respect to the observer. This set of coordinates is invariant with
respect to rotation and more straightforward to use in data analysis.
There are three reasons for differences between ``plane-parallel'' and
``wide-angle'' predictions:
\begin{description}
\item{(a)} The ``wide-angle'' correlation function $\xi(r,\mu,\alpha)$ depends on 
  $\alpha$, while the ``plane-parallel'' one doesn't;
\item{(b)} The coefficients $B^{\ell_1,\ell_2,\ell}$ depend on the density of
  galaxies $n(z)$ as a function of redshift. This implies that the RSD
  effect will depend on the spatial distribution of observed galaxies;
\item{(c)} The distribution of galaxies in $\mu$ will be non-trivial,
  with some values of $\mu$ not permitted for non-zero $\alpha$. As a
  consequence, we will not be able to measure pure Legendre momenta of
  the correlation function, but instead will use weighted integrals
  and biased momenta.
\end{description}
In \citet{raccanelli10} we used simulations to demonstrate these
effects, showing that they should be carefully taken into
consideration in order to fit the measured wide-angle correlation
function. In the following we describe (a) and (b) as wide-angle effects,
whereas (c) is termed the ``$\mu$-distribution'' as it could be applied to
plane-parallel and wide-angle theory of individual line-of-sight. 

\begin{figure}
  \includegraphics[height=160mm,width=80mm]{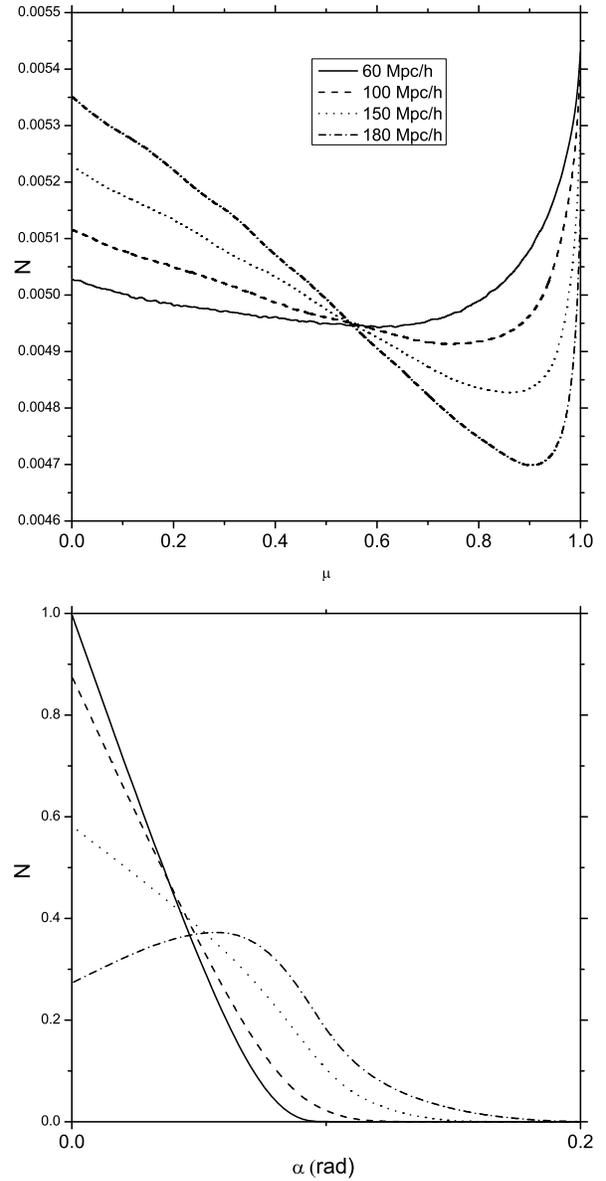}
  \caption{Normalised distribution of pairs in SDSS DR7 LRG catalog as
  a function of $\mu$ and $\alpha$ at different scales. Top panel shows
  distribution of $\mu$ for a fixed $r$ summed over all values of $\alpha$.
  Bottom panel shows distribution of $\alpha$ for fixed $r$ summed over all
  values of $\mu$.}   
    \label{fig:pdist}
\end{figure}

Allowing for full distribution of galaxy pairs, the estimates of Legendre momenta given
by Eq.~(\ref{eq:ls_ell}) correspond to
\begin{equation}
  \label{eq:wath}
  \xi_\ell(r)=\displaystyle\int \xi
   (r,\mu,\alpha)W(r,\mu,\alpha)P_\ell(\mu)d\mu d\alpha,
\end{equation}
where $\xi(r,\mu,\alpha)$ is given by either the wide-angle formula in
Eq.~(\ref{eq:trip}) or its plane-parallel equivalent computed from
Eq.~(\ref{eq:kaiser}).
$W(r,\mu,\alpha)$ is a weight factor that gives the relative number of pairs in
a survey that form angles $\mu$ and $\alpha$ for a given scale $r$. The weight
factor $W$ is normalised so
that
\begin{equation}  \label{eq:Wnorm}
  \displaystyle\int W(r,\mu,\alpha) d\mu d\alpha = 1
\end{equation}
for all scales $r$.  Ignoring (a) corresponds to setting
$W(r,\mu,\alpha\neq0)=0$ and ignoring (c) corresponds to setting
$W(r,\mu,\alpha)=1$.

In practice, $W(r,\mu,\alpha)$ weights can be computed from the random catalog;
they will be given by properly normalised $RR(r,\mu,\alpha)$ number counts. In
Sec.~\ref{ssec:zdist} we showed that uncertainties in radial distribution (that
translate into uncertainties in $RR$ counts and consequentially
into uncertainties in the $W$ weights) significantly affect measurements of
correlation function. This uncertainty will also bias our modelling of
$\mu$-distribution effects. This effect is of higher order and we do not
investigate it further in our paper.

Fig.~\ref{fig:pdist} shows the normalized distribution of pairs in $\mu$ and
$\alpha$ for different scales for the SDSS DR7 LRG catalog. When the $\alpha$
distribution tends towards a delta function centred at $\alpha=0$, the
wide-angle effects (a) becomes negligible. When the distribution in $\mu$ tends
towards a uniform one, (c) becomes negligible. In general, the relative
importance of the wide-angle effects on the measured correlation function
depends on the geometry of the survey, its redshift range and what scales are
considered. The effect is stronger for lower redshifts and becomes increasingly
important on larger scales. Top panel of Fig.~\ref{fig:pdist} implies that for
SDSS DR7 observed geometry it is easier to fit galaxy pairs across and along the
line of sight rather than for angles in between, more so for larger scales.  In
Sec.~\ref{ssec:mocktests} we will show that for the SDSS DR7 geometry, the
difference due to (a) and (b) in the list above are much smaller than
statistical errors and can be safely ignored even for scales as large as
$200\mpcoh$. The differences due to a nontrivial $\mu$-distribution (item
(c) in the list above) are larger than wide-angle effects but small compared to
current statistical errors. They are, however of order of few percent at larger
scales and will be important for future surveys. 

\subsection{Nonlinear effects}  \label{ssec:nl}

The following nonlinear effects, if they are comparable to the
measurement errors, can make Eq.~(\ref{eq:kaiser}) unsuitable for
analysing RSD data
\begin{description}
\item{(a)} Nonlinear contributions to the relationship between matter
  and galaxy overdensities $\delta_{\rm g}^r=h(\delta_{\rm m}^r)$, where
  $h$ is an arbitrary function.
\item{(b)} The relationship between the velocity divergence and matter
  overdensities $\theta_{\rm g} = -f \delta_{\rm m}$ relies on
  scale-independent linear growth coupled with the continuity
  equation. Also the galaxy velocity divergence field must be an unbiased tracer
  of the matter velocity divergence field, i.e., $\theta_{\rm g} = \theta_{\rm
  m}$. This formula will break down if these conditions are not
  met.
\item{(c)} The matter power spectrum itself goes non-linear, because of the
  scale-dependent non-linear growth on smaller scales.
\item{(d)} The real-space to redshift-space mapping includes higher
  order terms involving $\delta_{\rm g}^r$ and $\theta_{\rm g}$.\footnote{This
  refers to the nonlinear effects in the large-scale coherent motions and not to
  FOG nonlinearities due to virialization of galaxy velocities within halos
  which will be discussed in Sec.~\ref{ssec:fog}.}
\end{description}
In the following, we are only interested in the signal on large-scales where
linear theory should be strongest. We therefore assume that all non-linear
effects are small except for (c), where we allow the overall power spectrum
shape to deviate from the linear form (but see, \citet{reid11}). We use mock
catalogues in Section~\ref{ssec:mocktests} to confirm the validity of this
assumption.

In order to approximate the non-linear power spectrum, we adopt a
two-component model, which splits $P(k)$ into a ``smooth'' part that
describes the overall shape and a ``wiggled'' part that describes the
Baryon Acoustic Oscillations (BAO),
\begin{equation}  \label{eq:pwiggled}
  P_{\rm bao}(k,\mu) = P_{\rm full}(k,\mu)-P_{\rm smooth}(k,\mu)
\end{equation}
The ``smooth'' part is defined by taking some reasonably spaced points $k_i$
and then interpolating the linear power spectrum values between those nodes
using a bi-cubic spline interpolation routine \citep{press92}. In this work we
use $k_i$ spacing similar to \citealt{percival10}; we place nodes at
$k=0.001$, $k=0.25$ and $k=0.25 + 0.05n$ where $n$ is large enough for the
purposes of recovering the correlation function by the means of a Fourier
transform.\footnote{When making actual Fourier transform we are using much denser set
of nodes compared to the ones used to define ``smooth'' power-spectrum to make
sure that the noise introduced by discreetness of Fourier transform is small.}

The primary non-linear effect on the BAO component of the power
spectrum is a damping on small scales, which can be well approximated
by a Gaussian smoothing \citep{bharadwaj96,crocce06,crocce08,
  eisenstein07,matsubara08a,matsubara08b}
\begin{eqnarray}  \label{eq:nlbao}
  P^{\rm nl}_{\rm bao}(k,\mu) &= &P^{\rm lin}_{\rm bao}(k,\mu) \times \nonumber \\
    &&\exp\left(-k^2\left[\frac{(1-\mu^2)\Sigma_\bot}{2}
        +\frac{\mu^2\Sigma_{||}}{2}\right]\right),
\end{eqnarray}
where $\Sigma_\bot=\Sigma_0G$ and
$\Sigma_{||}=\Sigma_0G(1+f)$. $\Sigma_0$ is a constant
phenomenologically describing the nonlinear diffusion of the BAO peak due to
nonlinear evolution. From N-body simulations its numerical value is of order
10$\mpcoh$ and seems to depend linearly on $\sigma_8$ but only weakly on $k$ and
cosmological parameters. 

Next order non-linear effect results in a tilt of correlation function on large
scales just before the BAO peak \citep[For details see,][]{sanchez09}. We do not
consider this and other higher order terms in our computations. Robust data
analysis of future high quality measurements should also include a modelling of
this small scale nonlinear effects.

\subsection{Fingers of god effect}  \label{ssec:fog}

Within dark matter haloes the peculiar velocities of galaxies are highly
non-linear. These velocities can induce RSD that are larger than the real-space
distance between galaxies within the halo. This gives rise to the observed
fingers of god (FOG) effect -- strong elongation of structures along the line of
sight \citep{jackson72}. The FOG effect gives a sharp reduction of the power
spectrum on small scales compared to the predictions of the linear model, and is
usually modeled by multiplying the linear power-spectrum by a function
$F(\sigma_{ v},k,\mu)$, where $\sigma_{v}$ is the average velocity
dispersion of galaxies within the relevant haloes. The function $F$ is chosen so
that it is small on small scales and approaches unity for scales larger than
$1/\sigma_{ v}$. The two most frequently used functions are
\citep[e.g.][]{cole95,PeaDod96}
\begin{eqnarray}
  F_{\rm Lorentzian}(k,\mu^2) &=& \left[1+(k\sigma_{v}\mu)^2\right]^{-1} 
    \label{eq:F_exp}, \\
    F_{\rm Gaussian}(k,\mu^2) &=& \exp\left[-(k\sigma_{v}\mu)^2\right]
    \label{eq:F_gauss}.
\end{eqnarray}
Note that this model is constructed by a rather ad-hoc splicing of the FOG
signal together with the linear model and ignores the scale-dependence of the
mapping between real and redshift-space separations \citep[][and references
therein]{Fis95,scoccimarro04}. In addition, the exact form of $F(k,\mu^2)$, and
the value of $\sigma_{v}$ is strongly dependent on the galaxy population
\citep{Jing04,Li07}. 

The Gaussian smoothing in Eq.~(\ref{eq:nlbao}), amongst other nonlinear effects,
also accounts for the damping due to random velocities described by
Eqs.~(\ref{eq:F_exp})--(\ref{eq:F_gauss}). In our analysis we will use a model
given by Eq.~(\ref{eq:nlbao}) that partially includes the FOG effect on large
scales and will ignore FOG effects on small scales.

\subsection{Degeneracy with Alcock-Paczynski effect}
\label{ssec:ap}

The positions of galaxies in our catalog are given in terms of the
angular positions and redshifts. To convert angular and redshift
separations into physical distances the angular and radial distances
as functions of redshift are required. Those functions depend on the
adopted cosmological model. We perform our pair count assuming a
fiducial, spatially-flat cosmology with $\Omega_{\rm m}=0.25$. If the
real cosmology is significantly different from the fiducial one, this difference
will introduce additional anisotropies in the correlation function through the
Alcock-Paczynski (AP) effect \citep{alcock79}. This can significantly bias the
measurements of growth \citep{ballinger96,simpson10,samushia10}.

In the presence of the AP effect the redshift-space power-spectrum is
\begin{eqnarray}
  P^{\rm s}(k',\mu',\alpha_\bot,\alpha_{||},{\bf p})&=&
  \left(b+\frac{\mu'^{2}f}{F^2+\mu^2(1-F^2)}\right)^2\alpha_{\bot}^{-2}\alpha_{||}^{-1}\times \nonumber \\
  &P^{\rm r}&\left(\frac{k'}{\alpha_\bot}\sqrt{1+\mu^{'2}\left(\frac{1}{F^2}-1\right)}\right),
  \label{eq:rsdap}
\end{eqnarray}
\noindent
where ${\bf p}$ are standard cosmological parameters determining the shape of
the real-space power-spectrum, $k'$ and $\mu'$ are the observed wavevector and
angle, related to the real quantities by
\begin{eqnarray}
  k'_{||}&=&\alpha_{||}k_{||},\\
  \label{eq:kprime1}
  k'_\bot&=&\alpha_\bot k_\bot,\\
  \label{eq:kprime2}
  \mu'&=&\frac{k_{||}'}{\sqrt{k_{||}'^2+k_\bot'^2}},
  \label{eq:kprime3}
\end{eqnarray}
\noindent
the $\alpha_{||}$ and $\alpha_\bot$ are the ratios of angular and radial
distances between fiducial and real cosmologies
\begin{eqnarray}
  \alpha_{||}&=&\frac{H^{\rm fid}}{H^{\rm real}},\\
  \label{eq:alpha1}
  \alpha_\bot&=&\frac{D^{\rm real}}{D^{\rm fid}},
  \label{eq:alpha2}
\end{eqnarray}
\noindent
and $F=\alpha_{||}/\alpha_\bot$.

Ignoring the AP effect is equivalent to assuming that $\alpha$ factors
are equal to unity in Eq.~(\ref{eq:rsdap}). This assumption can bias 
estimates of growth parameters and their uncertainties.

We estimate to magnitude of this effect for our analysis using Fisher matrix
method. We compute a Fisher matrix for the SDSS-II like survey following
\cite{samushia10}. This Fisher matrix is an optimistic estimate of the inverse
covariance matrix on the parameters $b$, $f$, $\alpha_{||}$, $\alpha_\bot$ and
${\bf p}$.  The covariance matrix is an inverse of this fisher matrix.  Ignoring
the Alcock-Paczynski effect is equivalent to removing rows and columns
corresponding to $\alpha_{||}$ and $\alpha_\bot$ first, as if they were
perfectly known, and only then inverting the fisher matrix to get covariance of
$b$ and $f$. The more accurate approach is to invert the Fisher matrix directly
without assuming that the $\alpha$-s are known.

In our data analysis, we will apply a prior based on the WMAP and SNIa data on
the background geometry of the Universe (see Sec.~\ref{sec:meas}). To reflect
this in our Fisher matrix computations we first add this prior to the Fisher
matrix elements corresponding to $\alpha_{||}$ and $\alpha_\bot$ and only then
invert the whole matrix to get covariances on $b$ and $f$. We compare the result
with the resulting covariances when the AP effect is ignored. 

To do this we use MCMC chains corresponding to WMAP and SNIa joint constraints
on spatially-flat WCDM Universe from WMAP LAMBDA
website\footnote{\url{http://lambda.gsfc.nasa.gov/product/map/dr4/params/wcdm_sz_lens_wmap7_snconst.cfm}}
to estimate Fisher matrix of parameters $\Omega_{\rm m}$, $w_0$ and $H_0$. The
actual priors are $\Omega_{\rm m}=0.276\pm0.020$, $w_0=-0.969\pm0.054$, $w_a=0$,
$\Omega_{\rm k}=0$ while the errorbars are Gaussian and slightly correlated.  We
transform this into a Fisher matrix on $\alpha_{||}$ and $\alpha_\bot$.

\begin{figure}
  \includegraphics[height=80mm,width=80mm]{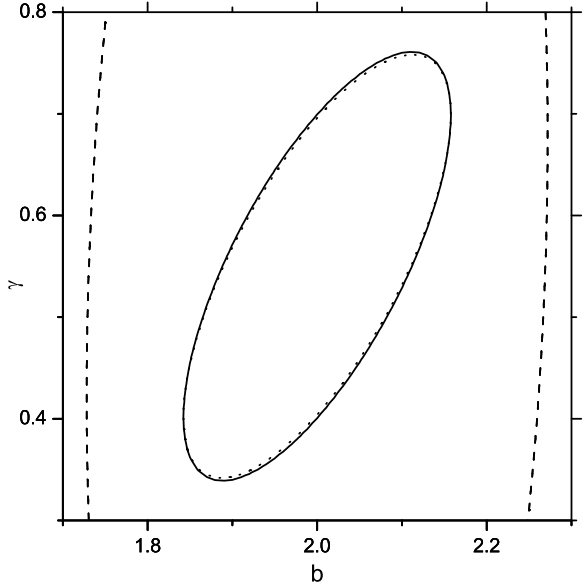}
  \caption{Fisher matrix predictions of correlated constraints on parameters $b$
  and $\gamma$ from a SDSS-II like survey. The dashed contours correspond to the
  most general case when the AP effect is not ignored, the dotted lines correspond
  to the case when the AP effect is ignored and the solid lines correspond to the
  case when the AP effect is ignored but a strong prior is put on the background
  cosmology. The solid and dotted lines are almost indistinguishable by eye on
  this plot.}
  \label{fig:apbg}
\end{figure}

\begin{figure}
  \includegraphics[height=80mm,width=80mm]{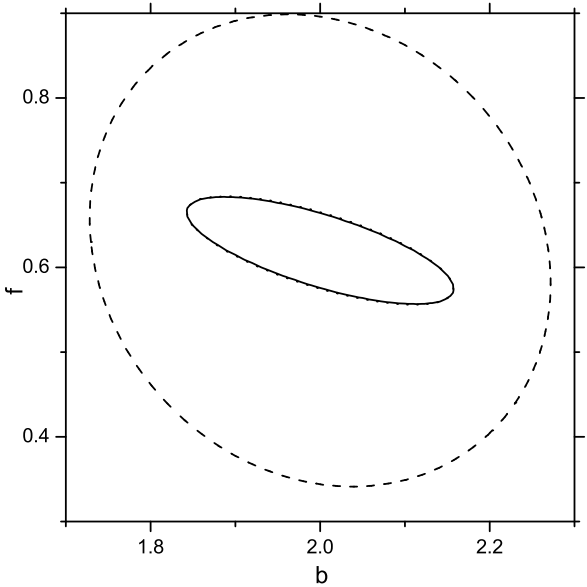}
  \caption{Fisher matrix predictions of correlated constraints on parameters $b$
  and $f$ from a SDSS-II like survey. The dashed contours correspond to the
  most general case when the AP effect is not ignored, the dotted lines correspond
  to the case when the AP effect is ignored and the solid lines correspond to the
  case when the AP effect is ignored but a strong prior is put on the background
  cosmology. The solid and dotted lines are almost indistinguishable by eye on
  this plot.}
  \label{fig:apbf}
\end{figure}

Figures \ref{fig:apbg} and \ref{fig:apbf} show the effects of AP
on the measurements of different growth parameters (for the
description of parameters $\gamma$ and $f$ see
Sec.~\ref{sec:meas}). These figures show that in general ignoring the
AP effect results in gross underestimation of the error bars. Real
uncertainties on $\gamma$ and $f$ are few times larger then what we
would get when ignoring AP. After applying the strong prior on the
background expansion, however, almost all of this degeneracy is
removed and the uncertainties in the measurements of growth and bias
are almost identical to the case with no AP, consistent with the work
of \cite{samushia10}, which showed the importance of model assumptions
on this measurement. We conclude that, for the models we test, the effects of the
degeneracy between RSD and AP on the error bars of our measurements
are very small and can be safely ignored provided we adopt joint WMAP and SNIa
priors.

\subsection{Relative importance of different effects}  \label{ssec:modsum}

\begin{figure}
  \includegraphics[height=160mm,width=80mm]{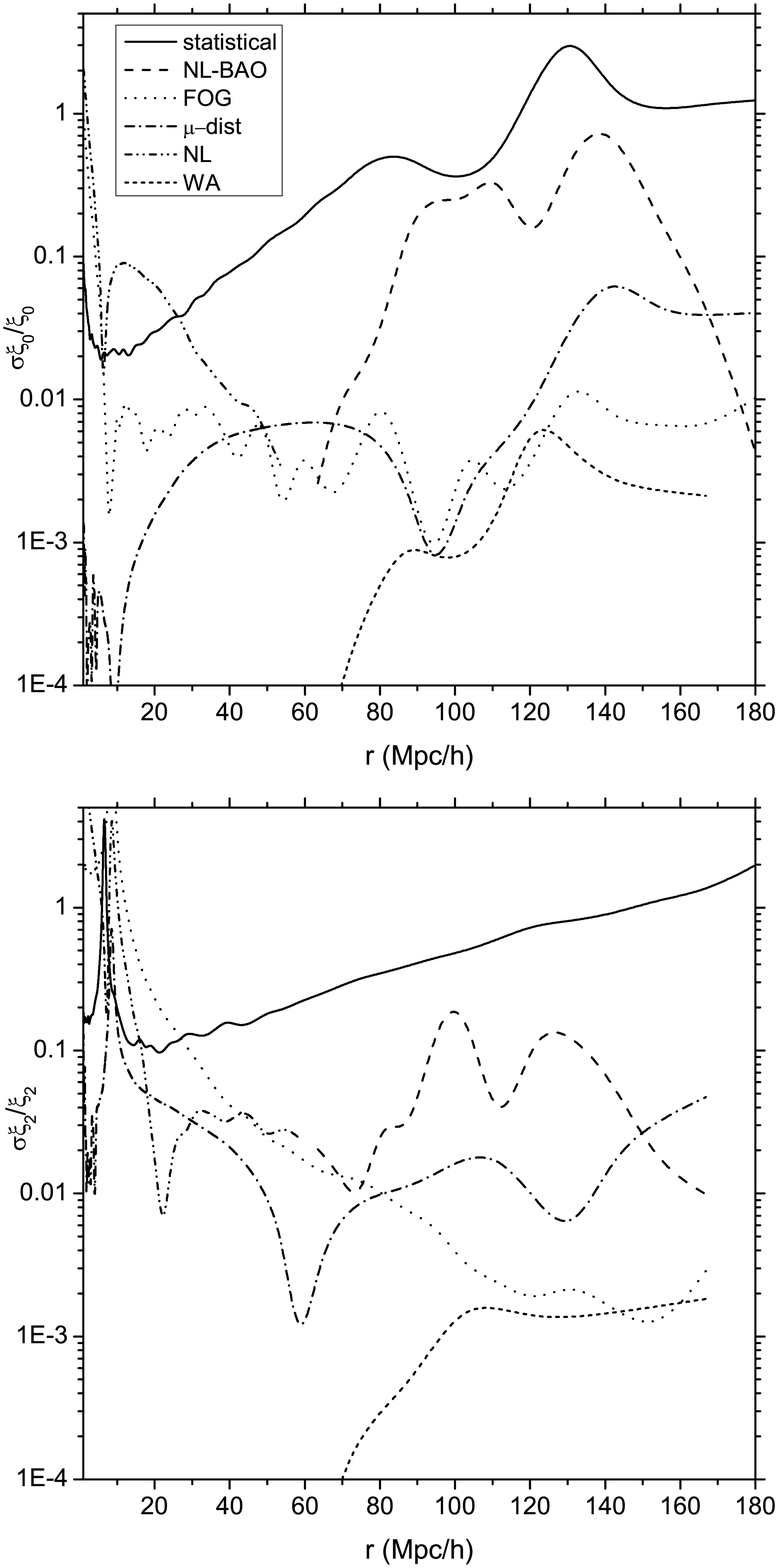}
  \caption{Relative impact of nonlinear and wide-angle effects compared to the
    statistical errors on the measurements of $\xi_\ell(r)$ from SDSS DR7 data.}
  \label{fig:xisum}
\end{figure}

\begin{figure}
  \includegraphics[height=80mm,width=80mm]{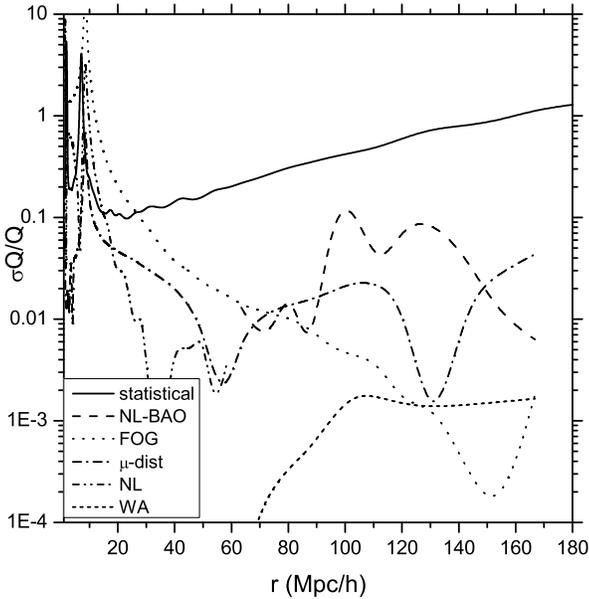}
  \caption{Relative impact of nonlinear and wide-angle effects compared to the
    statistical errors on the measurements of $Q(r)$ from SDSS DR7 data.}
  \label{fig:Qsum}
\end{figure}

We now consider the relative importance of the modifications to the linear
plane-parallel model, described in previous subsections, as a function of scale.
We have assumed a spatially-flat $\Lambda$CDM model with $\Omega_{\rm m}=0.25$
throughout. Figs.~\ref{fig:xisum} and~\ref{fig:Qsum} compare statistical errors
on measurements of $\xi_\ell(r)$ and $Q(r)$ (for the details of how these
statistical errors are estimated see Sec.~\ref{ssec:mocktests}) with the
differences in best-fit theoretical models calculated with or without the
modifications considered above.

To estimate the impact of different systematics we first compute a theoretical
correlation function for our fiducial model $\xi({\bf r})^{\rm full}$ including
all effects. We compute linear $\xi({\bf r})$ using {\sc camb} \citep{camb}. Then we
recompute the same correlation function by ignoring each of the systematic
effects in turn to see by how much this changes our theoretical estimates at
different scales.  Nonlinear diffusion of the BAO peak is modelled with the
Eq.~(\ref{eq:nlbao}) with $\Sigma_0=8\mpcoh$;\footnote{The value is consistent
with the estimate of $\Sigma_0$ for a real-space power-spectrum on low redshifts
and at large scale computed from N-body simulations in a standard $\Lambda$CDM
\citep{eisenstein07}.} the FOG effect with the Gaussian damping function of
Eq.~(\ref{eq:F_gauss}) with $\sigma_{ v}=3.5\mpcoh$;\footnote{This value is
consistent with recent measurements from \citet{song10}; slightly lower than the
estimates in \citet{cabre09}.} the effect of $\mu$ distribution is studied first
by using the real distribution of angles in SDSS geometry for $W$ and then
assuming it to have a uniform probability over all angles; the magnitude of the
effect to small scale nonlinearities is calculated by comparing correlation
functions computed from linear power-spectrum to the
nonlinear one (computed using the HALOFIT fitting formula of \citealt{smith03});
the wide-angle effects are estimated by substituting the full wide-angle
correlation function by a two-dimensional plane-parallel approximation.

The results are shown on Figs.~\ref{fig:xisum} and \ref{fig:Qsum}.  The FOG
effect and corrections to the shape of the correlation function due to nonlinear
growth of structure are only important on smaller scales and are lower than the
measurement errors on the scales larger than $20\mpcoh$ for the monopole and
scales larger than $30\mpcoh$ for the quadrupole. It should be noted that the
FOG induced relative errors as shown on Figs.~\ref{fig:xisum} and \ref{fig:Qsum}
correspond to the difference between using the anisotropic Gaussian dumping of
Eq.~(\ref{eq:F_gauss}) and ignoring the effect altogether, for a wrong but
reasonable FOG modelling the induced systematic errors will be smaller. The wide
angle effects appear at the scales of about $70\mpcoh$, as anticipated based on
Fig.~\ref{fig:pdist}, but are less then 1 percent even on the scales as large as
$200\mpcoh$. The effects of nonlinear BAO diffusion become comparable to the
error-bars on scales between $80\mpcoh$ and $200\mpcoh$ and should be taken into
account to get accurate theoretical predictions. The effects due to a non-flat
$\mu$-distribution are next in order of importance after nonlinearities on large
scales. For current data sets these effects are small compared to the
statistical errors, but they will become important for future surveys. In what
follows we will therefore only use the data on scales between $30\mpcoh$ and
$200\mpcoh$, and will ignore the nonlinear FOG, nonlinear growth other than BAO
diffusion and wide-angle effects, but will take into account the effects of
nonlinear BAO diffusion and the $\mu$ distribution.

In Sec.~\ref{ssec:mocktests} we test the applicability of our model on the mock
catalogs. The measurements of mean $\xi_\ell$ from the mocks have statistical
error bars that are approximately nine times smaller compared to the SDSS data.
To fit the mock measurements accurately we will also have to take into account
the FOG effect.

To summarise, our theoretical model of the correlation function will
be given by
\begin{equation}  \label{eq:mod}
  \xi_\ell(r)^{\rm th}=\displaystyle\int \xi^{\rm
    th}(r,\mu,\alpha)W(r,\mu,\alpha)P_\ell(\mu)d\mu d\alpha,
\end{equation}
where the function $\xi^{\rm th}(r,\mu,\alpha)$ is computed by Fourier
transforming a power-spectrum given by formula in
Eq.~(\ref{eq:kaiser}). We will model the real-space power-spectrum on
the right hand side of Eq.~(\ref{eq:kaiser}) as a linear
power-spectrum damped with a Gaussian function of Eq.~(\ref{eq:nlbao})
to account for nonlinear diffusion of the BAO peak. The other effects are
considered negligible for the SDSS data on these scales.

\subsection{Testing RSD models with mock catalogs}  \label{ssec:mocktests}

To test our analysis of the effects that have to be taken into account
to analyse RSD in SDSS DR7 data, and to estimate the statistical
errors on our measurements (as shown on Fig.~\ref{fig:xim}), we use
galaxy catalogs from the Large Suite of Dark Matter Simulations
(LasDamas: \citealt{mcbride11})\footnote{\tt
  http://lss.phy.vanderbilt.edu/lasdamas/}. The LasDamas simulations
are designed to model the clustering of Sloan Digital Sky Survey
(SDSS) galaxies in a wide luminosity range and in the redshift range
$0.16<z<0.44$. The simulations are produced by placing artificial
galaxies inside dark matter halos using an HOD with parameters
measured from the respective SDSS galaxy samples. We use 80 ``Oriana''
catalogs that have exactly the same angular mask as the SDSS survey
and subsample them to match the redshift distribution of the Luminous
Red Galaxies (LRG) in our SDSS DR7 data set.  The LasDamas mocks have
insufficient galaxies at redshifts below $z<0.2$, as a result the mocks will
slightly overestimate the shot noise. We do not expect this to be important
since the effected region contains only a small fraction of the volume
available.

We apply exactly the same weighting to the mocks as to the real
catalog and compute zeroth, second and fourth Legendre momenta of the
redshift-space correlation function from them using
Eq.~(\ref{eq:ls_ell}). We also compute the normalised quadrupole
$Q(r)$ as given by Eq.~(\ref{eq:Qm}).

We estimate covariance matrices corresponding to the statistical
errors of our measurements, based on assuming that the Legendre
momenta are drawn from a multi-variate Gaussian distribution
\begin{equation}  \label{eq:cstat}
  {\bf C}^{\rm stat} = \frac{1}{79}\displaystyle\sum 
    \left[\hat{\bf X}(r_i)-{\bf \overline{X}}(r_i)\right]
    \left[\hat{\bf X}(r_j)-{\bf \overline{X}}(r_j)\right],
\end{equation}
where $\hat{\bf X}(r)$ is a vector of the measurements of $\xi_\ell$ at scale
$r$ for $\ell=0,2,4$ and ${\bf \overline{X}}$ is the mean value from all 80
mock catalogs.

\begin{figure}
  \includegraphics[height=210mm,width=80mm]{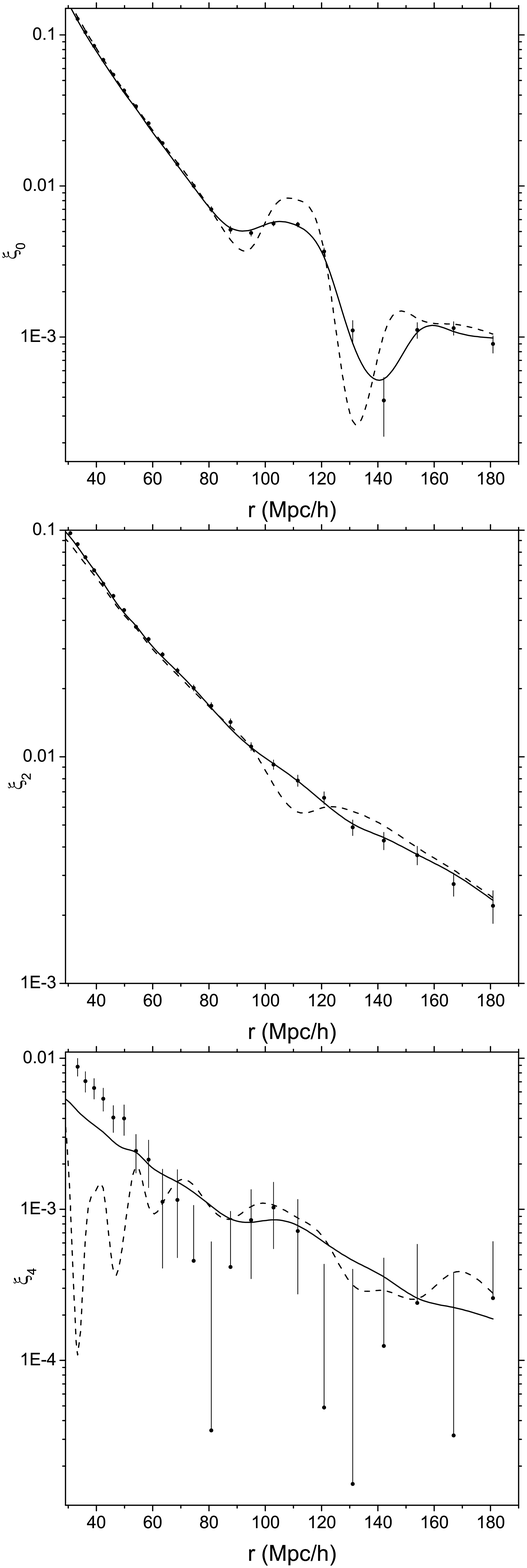}
  \caption{Measurements of mean $|\xi_\ell|$ from 80 LasDamas ``Oriana'' mocks in
  a redshift range $0.16<z<0.44$. Dashed line shows predictions of the Kaiser
  model (for the best fit values of $b\sigma_8$, $f\sigma_8$ and $\sigma_v$),
  while solid line corresponds to the theoretical predictions of the model with
  nonlinear BAO damping and non-flat $\mu$-distribution (for the best fit values
  of $b\sigma_8$, $f\sigma_8$, $\sigma_v$ and $\Sigma_0$). }
  \label{fig:ximock}
\end{figure}

The mean Legendre momenta measured from the LasDamas mocks are shown
in Fig.~\ref{fig:ximock}. The error-bars correspond to the square root
of the diagonal terms in the covariance matrix ${\bf C}^{\rm stat}/80$
and the lines show theoretical predictions computed making different
assumptions. Our theoretical predictions, with the parameters of the
simulations, provide a very good fit to the data. The bottom panel on
Fig.~\ref{fig:ximock} shows that the theoretical prediction underestimates
$\xi_4$ on scales smaller than 50$\mpcoh$. The fourth Legendre moment measures a
higher frequency $\mu$ dependence of correlation functions and therefore is more
sensitive to different systematic effects. Since we are not using $\xi_4(r)$ in
our fits we did not attempt to investigate this issue further. The measurements
are strongly positively correlated and the error-bars presented here reflect
only small part of the covariance matrix.

\begin{figure}
  \includegraphics[height=80mm,width=80mm]{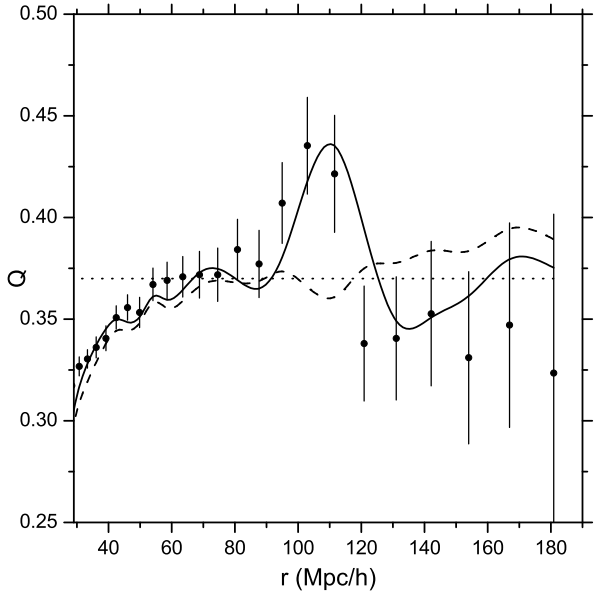}
  \caption{Measurements of mean $Q$ from 80 LasDamas ``Oriana'' mocks in a
  redshift range $0.16<z<0.44$. Dotted line shows theoretical predictions of
  Kaiser model, dashed line shows predictions of the Kaiser model with FOG
  corrections, modeled with a Gaussian function, and accounting for the fact
  that $Q$ is computed from a discrete sum, while solid line corresponds to the
  theoretical predictions of our model. Wiggles in solid and dashed lines are
  related to the fact that the integral in the definition of $Q$ is replaced by
  discrete sum and we expect similar wiggles in the data.}
  \label{fig:Qmock}
\end{figure}

Figure \ref{fig:Qmock} shows the measurements of $Q$ from LasDamas mocks with the
similar definition of the error-bars as Fig.~\ref{fig:ximock}. In the Kaiser
formalism $Q(r)$ is expected to be a straight line damped at smaller scales
because of FOG effects. In reality the measured $Q(r)$ will deviate from a
straight line even within the Kaiser model at larger scales since it is
estimated by a discrete sum in Eq.~(\ref{eq:Qm}) instead of continuous integral.
Figs.~\ref{fig:ximock} and~\ref{fig:Qmock} clearly show that the model adopted
in Sec.~\ref{ssec:modsum} can describe the measurements very well on all scales
between $30\mpcoh$ and $200\mpcoh$, while using the Kaiser formula without
modifications would fail to fit on scales around BAO peak and larger.

\section{Testing Cosmological Models}
\label{sec:meas}

Figure~\ref{fig:xim} shows that the signal to noise of the measured $\xi_4$ is
very small on all scales so that it can not be used to extract RSD information.
Consequently, for simplicity, we will not include measurements of $\xi_4$ in our
analysis. Recent studies have shown that including hexadecapole in the fit
improves errors on measured cosmological parameters \citep{taruya11,kazin11}. We
find similar improvement only if the geometrical information is measured from
the survey itself through the AP effect without imposing any external priors. If the
background geometry is fixed by strong external priors, as is the
case in our analysis, the difference in measurements of $b\sigma_8$ and
$f\sigma_8$ is very small.

The normalised quadrupole $Q$ by definition does not contain any extra
information compared to $\xi_0$ and $\xi_2$. Figure~\ref{fig:Qsum} shows that
our measurements of $Q$ are noisier than first two Legendre momenta on scales
larger than 50$\mpcoh$. The analysis of $Q$ measurements is in some way simpler,
because the normalised quadrupole, under some general assumptions,\footnote{This
does not hold, for example, when the $\mu$ distribution does not correspond to
an isotropic distribution of galaxy pairs. Nonlinear effects will also make $Q$
deviate from a straight line.} does not depend on the shape of the
power spectrum and only depends on the parameters describing the amplitude. The
drawback is that with $Q$ only a combination of growth and bias $\beta=f/b$ can
be measured, but not the two parameters individually, and Fig.~\ref{fig:Qmock}
shows that $Q$ is model dependent \cite[in detail see also,][]{tocchini11}.

In our analysis we will perform a joint fit to $\xi_0$ and $\xi_2$. We will
explicitly use a prior based on the joint constraints from WMAP and SNIa data to
deal with the ambiguity in the shape of the matter power-spectrum.  This will
allow us to measure bias and growth at the same time and to extract more
information from the correlation function. Fitting the first two even Legendre
momenta of correlation function is advantageous compared with fitting to the two dimensional
correlation function $\xi({\bf r})$ for two reasons: a pair of
one-dimensional functions $\xi_0$ and $\xi_2$ are easier to
visualise and work with and, as we showed above, they contain most of
the cosmologically relevant information anyway; also the measurement
errors on $\xi_\ell$ are more Gaussian, as we show in
Sec.~\ref{ssec:likinac} compared to the errors of $\xi({\bf r})$ and
therefore the reconstruction of the likelihood surfaces is more
robust.

Our theoretical model of Legendre momenta of the correlation function will
depend on a set of parameters $\bf p$ describing background
expansion of the Universe and a set of parameters ${\bf A}(z)$
describing the amplitude of the correlation function and its growth
with redshift. Each model will also depend on the phenomenological
parameter $\Sigma_0$ describing nonlinear diffusion of the BAO
peak. We will treat $\Sigma_0$ as a nuisance parameter and marginalise
over it with a uniform prior. For this reason, we do not include the
$\Sigma_0$ dependence of the likelihood in the equations given below.

For the background expansion we will assume that the Universe is well
described by a spatially-flat wCDM model composed of non-relativistic
matter with energy density $\Omega_{\rm m}$ some part of which is in
baryons with energy density $\Omega_{\rm b}$. The rest of the energy
density, in this model, is assumed to be in a smooth dark fluid with
the equation of state $w$. To complete the background model we need to
specify the expansion rate of the Universe at present $h=H_0/100$,
where $H_0$ is a Hubble parameter.

For the cosmological parameters describing the observed amplitude of
clustering we will make three different assumptions ranging from the
most specific model to more general assumptions.

For every theoretical model we compute a $\chi^2$ function
\begin{equation}  \label{eq:chi}
  \chi^2_{\rm tot}({\bf p},{\bf A})=
    [{\bf \hat{X}-X}(r_i)]
    {\bf C}_{\rm tot}
    [{\bf \hat{X}-X}(r_j)]^{\rm T},
\end{equation}
where ${\bf \hat{X}}(r)$ is a vector of the measured $\xi_0(r)$ and
$\xi_2(r)$, ${\bf X}(r)$ is the model to be tested, and the
total covariance matrix is given by Eq.~(\ref{eq:cstat}).
Assuming that the measurement errors are closed to Gaussian, the likelihood for
a given set of cosmological parameters given data will be
\begin{equation}
  \label{eq:lik}
  \mathcal{L}_{\rm tot}=\exp(-\chi^2_{\rm tot}/2).
\end{equation}

\subsection{Inaccuracies in the estimation of covariance matrix}
\label{ssec:covinac}

Estimating covariance matrices of galaxy two-point correlation function in
configuration space is a nontrivial task. Many different techniques have been
used before to tackle this issue, including internal procedures -- based only on
the observed data itself -- such as jacknife \citep{lucey79} and bootstrap
\citep{barrow84} methods; analytical estimates of the errors \citep{mo92};
Monte-Carlo sampling of random initial conditions and combination of analytical
methods and Monte-Carlo \citep{padmanabhan05}. Studying the
three-dimensional clustering on scales below 25$\mpcoh$, \citet{norberg09}
showed that internal methods do recover the principal components of the real
covariance matrix in a robust way but can not accurately reproduce the errors
themselves, usually overestimating them by as much as 40 percent.

Our statistical covariance matrices are estimated from the sample of 80 mock
catalogs. This number is far less then sufficient to accurately measure the
errors and correlations. \citet{cabre08} used a sample of 1000 mocks, in their
study of cross-correlation between the map of CMB temperature anisotropies and
large scale structure, and found that even with 200 simulations the error bars
could be underestimated by about 20 percent. 

One of the ways of reducing the effect of inaccurate covariance matrix
estimation is to find the eigenvectors of the normalized covariance matrix and
then only use the eigenmodes that have high signal to noise, since there
error estimates are expected to be more reliable. We do not attempt to do this
in our paper; this would remove large scale information which we are interested
in. 

We will use the full covariance matrix estimated from 80 mock catalogs as our
best guess to the real structure of the measurement errors. This is good enough
for the purposes of our current work.  As we will show below the errors on
current data are too big to result in tight constraints on cosmological
parameters and we apply our method to the real SDSS-II data to provide a ``proof
of concept''. Next generation of surveys, with significantly tighter error bars
on the measurements of correlation function, will require a more thorough
investigation of this issue.

\subsection{Inaccuracies in the posterior likelihood function}
\label{ssec:likinac}

Equation~(\ref{eq:lik}) represents a true likelihood function only if the
measurements of variables
${\bf X}$ that were used in computing $\chi^2$ have errors that are distributed as
multivariate Gaussian random variables. There are reasons to believe that
the errors on $\xi_0$, for instance, are not Gaussian \citep{norberg09}. To check
if the assumption of Gaussianity holds reasonably well for our measurements we
take the measurements of $\xi_0$ and $\xi_2$ at the different scales from all 80
mock LasDamas catalogs and construct normalized variables 
\begin{equation}
  \label{eq:normgauss}
  {\bf Y} = \frac{ \hat{\bf X}-{\bf \overline{X}}}{\sigma_{\bf X}},
\end{equation}
\noindent
where ${\bf \overline{X}}$ and ${\sigma_{\bf X}}$ are the average value and
dispersion computed from all 80 mocks. If the measured $\hat{\bf X}$
are Gaussian, ${\bf Y}$ should be distributed according to the normal
distribution with mean zero and variance one.

\begin{figure}
  \includegraphics[height=160mm,width=80mm]{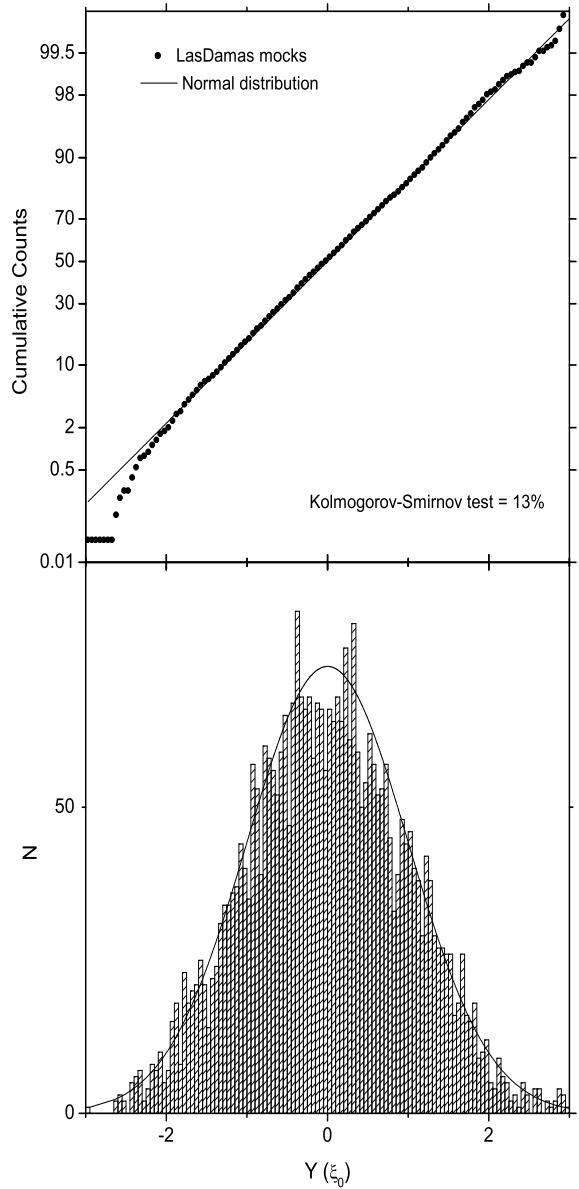}
  \caption{Histogram of the normalised scattering of $\xi_0$ measurements and
  their cumulative distribution from 80 mock
  LasDamas catalogs compared to the normal distribution with mean zero and unit
  variance.}
  \label{fig:ydist}
\end{figure}

\begin{figure}
  \includegraphics[height=160mm,width=80mm]{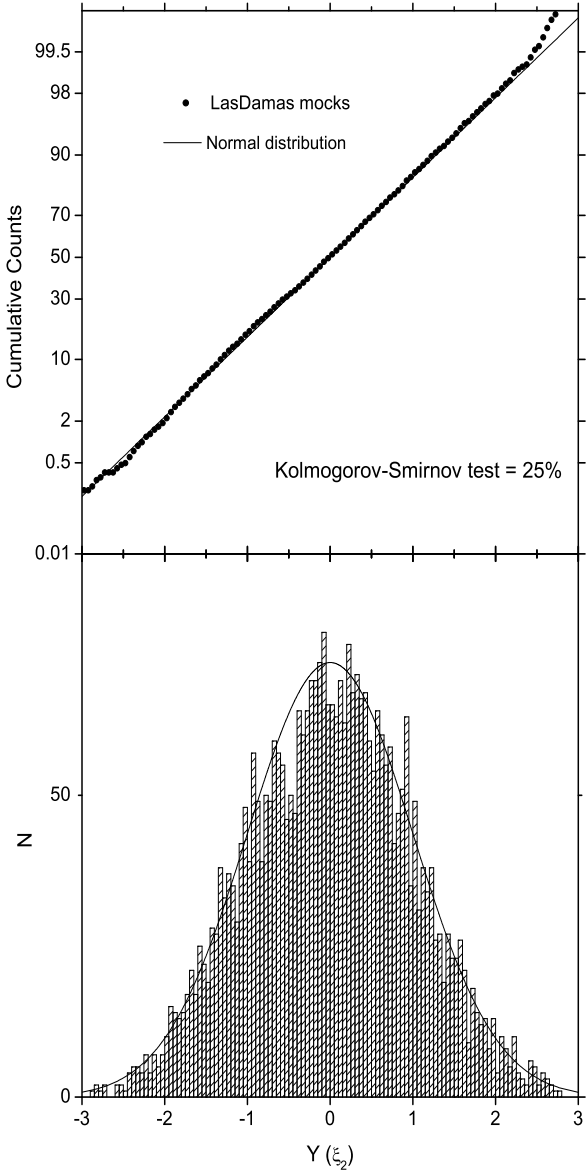}
  \caption{Histogram of the normalised scattering of $\xi_2$ measurements and
  their cumulative distribution from 80 mock
  LasDamas catalogs compared to the normal distribution with mean zero and unit
  variance.}
  \label{fig:ydist2}
\end{figure}

Figs.~\ref{fig:ydist} \&~\ref{fig:ydist2} shows the distribution of ${\bf Y}$ for the
measurements of $\xi_0$ and $\xi_2$. 

We perform a Kolmogorov-Smirnov test (for detailed discussion of
the test see, e.g., \citet{corder09}) to see if the empirical
distribution of $\xi_\ell$ measurements is consistent with the null hypothesis
that they are drawn from a Gaussian distribution. The Kolmogorov-Smirnov test
confirms that the distribution of $\xi_0$ is consistent with the null hypothesis
at 13 percent confidence level and $\xi_2$ is consistent at 25 percent
confidence level. In both cases the small possible deviations from Gaussian
cumulative distribution function reflect the fact that deviations above the mean
value are slightly more likely then deviations below the mean at the tails of
the distribution.

To check how the Gaussianity of $\xi_\ell$ measurement errors depends on the scale
we split $r$ range into two with $30\mpcoh<r_1<75\mpcoh$ and
$75\mpcoh<r_2<200\mpcoh$ and perform a similar Kolmogorov-Smirnov test on small
scale and large scale measurements separately. Our empirical distribution of
$\xi_0$ is more consistent with the assumption of Gaussianity on small scales.
For $r_1$ the KS test accepts the null hypothesis at 36 percent, while for $r_2$ the
null hypothesis is accepted at 8 percent. For the $\xi_2$ the trend is opposite KS
likelihood for $r_1$ is 5 percent, while for $r_2$ it is 17 percent. 

The variable $\log(1+\xi_0)$ is a slightly better fit to the assumption of
Gaussianity, with a KS likelihood of 23 percent over all scales.

For our purposes the variables $\xi_0$ and $\xi_2$ are close enough to
the Gaussian distributed variables and we conclude that the usage of
Eq.~(\ref{eq:chi}) is justified for computing likelihood surfaces and
confidence level intervals. 

\begin{figure}
  \includegraphics[height=160mm,width=80mm]{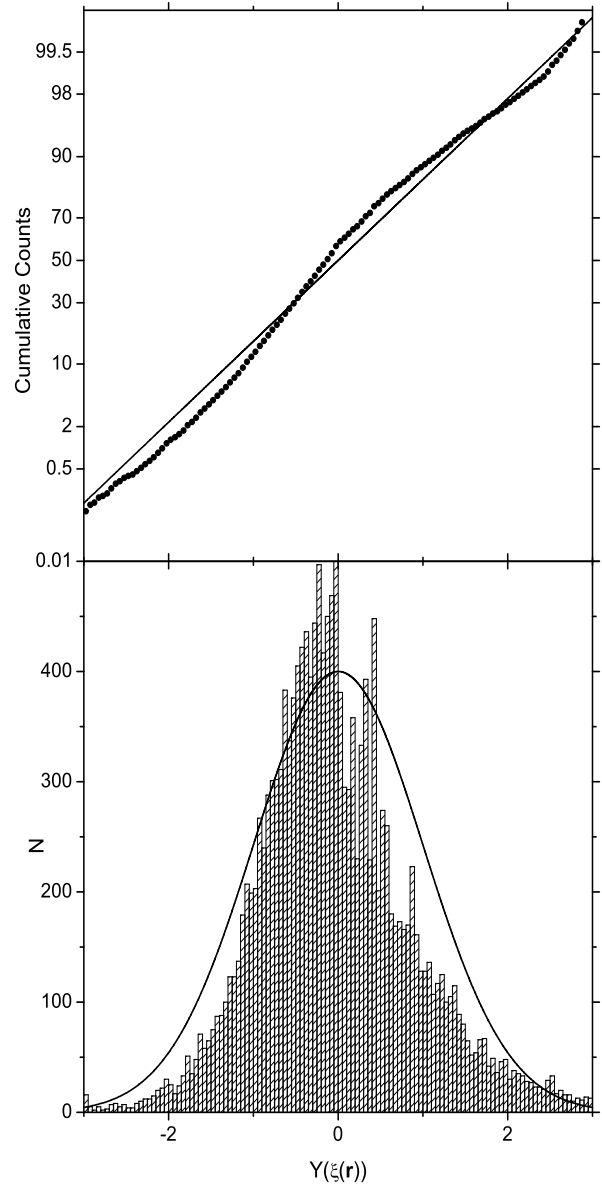}
  \caption{Histogram of the normalised scattering of $\xi(\sigma,\pi)$
  measurements from 80 mock LasDamas catalogs.}
  \label{fig:ydist3}
\end{figure}

The two dimensional redshift-space correlation function $\xi(\sigma,\pi)$, where
$\sigma$ and $\pi$ are along the line-of-sight and across the line-of-sight
separations, itself is often used in the analysis of RSD and BAO. We perform the
same check on the measurements of this two dimensional correlation function
between the scales of 30--60$\mpcoh$ histogramed in 40 two-dimensional bins. The
resulting histogram is shown on Fig.~\ref{fig:ydist3}. 

Figure.~\ref{fig:ydist3} shows that the distribution of measured
$\xi(\sigma,\pi)$ around its mean value, at least for the mock catalogs that we
use, are not Gaussian. The null-hypothesis that these random variables are drawn
from a Gaussian distribution is rejected at more than 50 percent significance level.
We conclude that if the $\chi^2$ and likelihood functions are defined as in
Eqs.~(\ref{eq:chi}) and (\ref{eq:lik}), the Legendre momenta are more
appropriate as variables compared to the two-dimensional correlation function.
This is not surprising since the measured Legendre momenta are the weighted sums
of the two-dimensional correlation function and will always tend to be more
Gaussian irrespective of the underlying distribution, thanks to the Central
Limit Theorem.

\subsection{Covariance matrix as a function of cosmological parameters}

Another systematic effect in estimating covariance matrix is the dependence of
${\bf C}^{\rm stat}$ on the cosmological model. Our estimates of the covariance
matrix are based on the mock catalogs that were created for a specific
cosmological model, namely a spatially-flat $\Lambda$CDM with $\Omega_{\rm
m}=0.25$, $\sigma_8=0.8$ assuming that the gravity is described well by GR. In
other cosmological models or different vales of parameters the intrinsic
scattering in the correlation function and therefore the covariance matrix will
be different. The scaling of ${\bf C}^{\rm stat}$ with cosmological parameters
is extremely difficult to model theoretically for nontrivial survey volumes. 

To estimate this effect we will again use the Fisher matrix calculations for an
SDSS-II like survey. We compute a Fisher matrix
$F(b,f,\alpha_{||},\alpha_\bot,{\bf p})$ for different values of cosmological
parameters and look at how the expected errors on the measurements of growth
scale. The uncertainties in the measurements of the power-spectrum can be
schematically divided into two parts: coming from the cosmic variance and from
the shot-noise. The shot-noise contribution depends on the total number of
galaxies and their distribution in the survey volume and is insensitive to the
underlying cosmological model. The cosmic variance component depends on the
parameters determining the overall amplitude of the power-spectrum $b$, $f$ and
$\sigma_8$, but is not very sensitive to the cosmological parameters describing
its shape ${\bf p}$.

\begin{figure}
  \includegraphics[height=80mm,width=80mm]{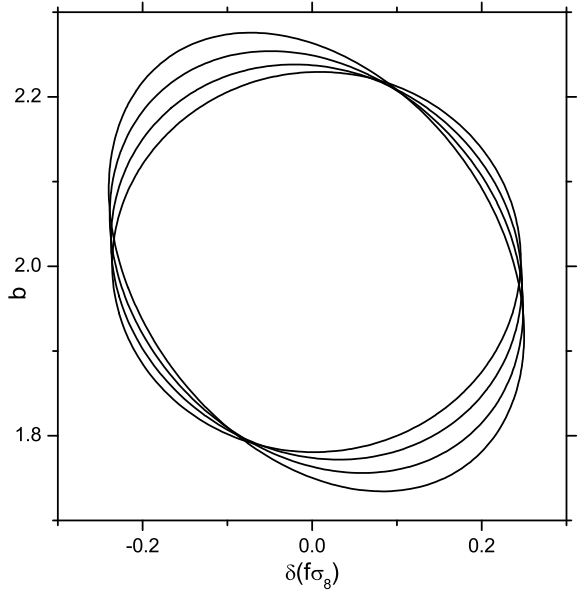}
  \caption{Fisher matrix predictions of the correlated errors of parameters $b$
  and $f\sigma_8$ from an SDSS-II like survey. Different contours correspond to
  the 1$\sigma$ confidence level ellipses for the fiducial cosmologies with
  $f\sigma_8$ equal to 0.2, 0.3, 0.4 and 0.5 from largest to smallest ellipse
  respectively. Bias is fixed to $b=2$. Horizontal tickmarks shows the deviation from
  fiducial value.}
  \label{fig:cp1}
\end{figure}

\begin{figure}
  \includegraphics[height=80mm,width=80mm]{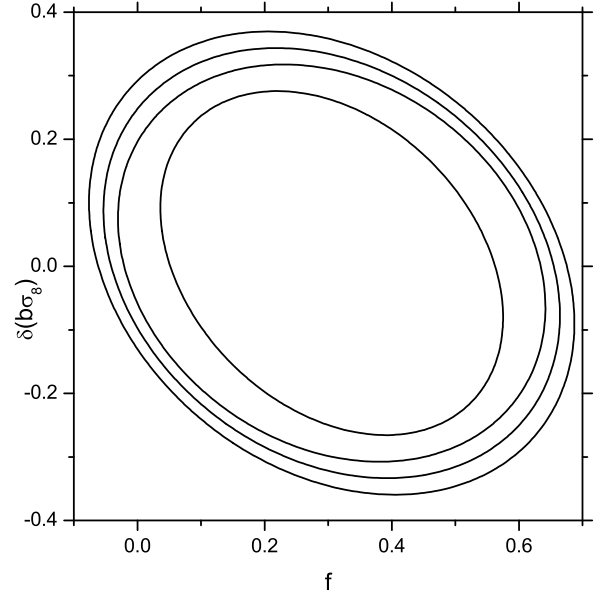}
  \caption{Fisher matrix predictions of the correlated errors of parameters
  $b\sigma_8$ and $f$ from an SDSS-II like survey. Different contours correspond
  to the 1$\sigma$ confidence level ellipses for the fiducial cosmologies with
  $b\sigma_8$ equal to 1.4, 1.5, 1.6 and 1.8 from outside to inside
  respectively. Growth rate is fixed to $f=0.45$. Vertical tickmarks shows the
  deviation from fiducial value.}
  \label{fig:cp2}
\end{figure}

We derive Fisher matrix errors on the measurements of the growth parameter
$f\sigma_8$ and bias $b\sigma_8$ for different fiducial values. This predictions
are shown on Figs.~\ref{fig:cp1} and \ref{fig:cp2}. Figure~\ref{fig:cp1} shows
the size of 1$\sigma$ ellipses for different fiducial values of $f\sigma_8$ when
$b=2$, while Fig.~\ref{fig:cp2} shows the same ellipses for different values of
$b\sigma_8$ when $f=0.45$. The relative change is small compared to the sizes of
the contours themselves. We conclude that this effect is relatively unimportant
for the range of values $f\sigma_8$ and $b\sigma_8$ allowed by our data
and ignore it in our analysis. 

For next generation surveys, however, the errors on the measurements of
growth will be significantly smaller and this effect will have to be taken into
account. This implies that deviations from the best-fit value towards stronger
clustering amplitude will be more likely then deviations of the same magnitude towards
weaker clustering amplitude.

\subsection{wCDM and General Relativity}
\label{ssec:gr}
In a specific cosmological model the growth rate will depend on the parameters
describing background geometry as well as the theory of gravity.
Assumptions about exact nature of this dependence bring in a very strong
theoretical priors that might affect the results of data analysis strongly. In
our work we will always make three separate assumptions about how the growth
rate depends on the background expansion from most restricting to almost premise
free.

First we will assume that General Relativity (GR) is the correct
theory of gravity. In this case the growth function $f^{\rm
  gr}(z)=f({\bf p},z)$ can be computed at every redshift from basic
cosmological parameters ${\bf p}$. The other two numbers that are
necessary to completely describe the amplitude of the correlation
function are the linear bias $b(z)$ and the overall amplitude of
clustering $\sigma_8(z=0)$. In this wCDM + GR model our cosmological
parameters of interest will be ${\bf p} = (\Omega_{\rm m},h,w)$ and
${\bf A}=(b_i(z),\sigma_8)$.

We will use the prior likelihood on ${\bf p}$ and $\sigma_8$ from the
WMAP7 measurements and SNIa data. We do this by using the relaxed
MonteCarlo Markov Chain of this joint data.\footnote{Available for
  download from \url{http://lambda.gsfc.nasa.gov}.} For every $b(z_i)$
we will go through the MCMC chain and for each value of ${\bf p}$ and
$\sigma_8$ compute the growth function $f(z)$ and then the theoretical
correlation function. Afterwards we will marginalize over ${\bf p}$ and
$\sigma_8$. This is equivalent to taking the following analytical
integral
\begin{equation}
  \label{eq:marg}
  \mathcal{L}^{\rm gr}(b(z_i)) = \displaystyle\int \mathcal{L}_{\rm tot}({\bf
  p},{\bf A})\mathcal{L}_{\rm prior}({\bf p},\sigma_8)d{\bf p}d\sigma_8,
\end{equation}
\noindent
where $\mathcal{L}_{\rm prior}$ is effectively given by the
MCMC. This will enable us to derive constraints on the linear bias
parameter $b(z)$ in two redshift bins.

\subsection{$\gamma$ parametrization of growth}
\label{ssec:gamma}

For our second model we will consider the $\gamma$ parametrization of growth
\citep{linder05}. In this model the growth function is assumed to depend on
parameters ${\bf p}$ as
\begin{equation}
  \label{eq:fg}
  f(z) = \left(\frac{\Omega_{\rm m}(1+z)^3}
    {\Omega_{\rm m}(1+z)^3+(1-\Omega_{\rm m})(1+z)^{-3(1+w)}}\right)^\gamma,
\end{equation}
where $\gamma$ is a redshift and scale independent number.  For
$\gamma=0.55$, Eq.~(\ref{eq:fg}) gives numerical results that are very
close to the predictions of GR. If $\gamma$ is larger than 0.55 the
growth is weaker compared to GR and vice versa.

In this model ${\bf A}=(b(z_i),\gamma,\sigma_8)$. We will assume that
the shape of the correlation function can still be accurately modelled
by wCDM predictions and will use WMAP + SNIa MCMC to marginalize over
${\bf p}$ and $\sigma_8$ as in Sec.~\ref{ssec:gr}.

To use the WMAP prior on $\sigma_8$ for $\gamma$ parametrization we
have to take into the account the fact that $\sigma_8$ is not directly
measured by CMB experiments. The measured quantity is $\sigma_8$ at
the last scattering surface with $z\approx1100$ and then the
$\sigma_8(z=0)$ is inferred by rescaling
\begin{equation}
  \label{eq:signow}
  \sigma_8(z=0) = \sigma_8(z=1100)\frac{G(z=0)}{G(z=1100)},
\end{equation}
\noindent
where $G(z)$ is the growth factor assuming GR.

To make the priors on $\sigma_8$ as given by WMAP MCMC chain
consistent with our assumption that the growth of structure is
modified as in Eq.~(\ref{eq:fg}) we rescale the values of $\sigma_8$ in
WMAP MCMC chain by
\begin{equation}
  \label{ewq:sigmod}
  \sigma_8(z=0)^\gamma = \sigma_8(z=0)^{\rm GR}\frac{G(z=1100)^{\rm
  GR}}{G(z=1100)^\gamma}\frac{G(z=0)^\gamma}{G(z=0)^{\rm GR}},
\end{equation}
\noindent
where $G^\gamma$ and $G^{\rm GR}$ are the growth functions computed in
$\gamma$ parametrization and GR respectively. After marginalization we
will get a posterior likelihood function
$\mathcal{L}^{\gamma}(b(z_i),\gamma)$. The measurements of $\gamma$ in
general will be correlated with the measurements of bias.

\subsection{Free growth}
\label{ssec:freegrowth}

In the last model we will not make any assumptions about the
relationship between $f$ and ${\bf p}$ and will treat $f(z_i)$ in each
redshift bin as a free parameter. In this model three parameters
describing the amplitude $f$, $b$ and $\sigma_8$ are degenerate and
only two combinations of them can be measured independently. We will
choose these combinations to be $b(z_i)\sigma_8(z_i)$ and
$f(z_i)\sigma_8(z_i)$.

We will assume again that only the growth of the perturbations is
different from the GR case and the overall shape of the correlation
function can still be model by wCDM model. If we make this assumption
we can use the same chains to marginalize over $\bf p$ so that we are
left with the posterior likelihood function $\mathcal{L}^{\rm
  fg}(b(z_i)\sigma_8(z_i),f(z_i)\sigma_8(z_i))$.

\section{Results and Discussion}
\label{sec:res}

\begin{center}
  \begin{table*}
    \begin{tabular}{|l|c|c|c|c|}
    \hline
     Model & Variable & Scales less than 60 ${\rm Mpc}/h$ & Scales up to 200 ${\rm Mpc}/h$ & ``Standard'' model expectation\\ 
    \hline
    \multirow{2}{*}{wCDM} & $b(z_1)\sigma_8(z_1)$ & 1.4216 $\pm$ 0.0724 & 1.3890 $\pm$ 0.0448 & \\ 
               & $b(z_2)\sigma_8(z_2)$ & 1.4053 $\pm$ 0.0582 & 1.50565 $\pm$ 0.0352 & \\ 
      \hline
        \multirow{3}{*}{$\gamma$} & $b(z_1)\sigma_8(z_1)$ & 1.4641 $\pm$ 0.0790 & 1.4156 $\pm$ 0.0491 & \\ 
                              & $b(z_2)\sigma_8(z_2)$ & 1.4641 $\pm$ 0.0703 & 1.5107 $\pm$ 0.0395 & \\ 
                            & $\gamma$ & 0.7366 $\pm$ 0.1638 & 0.5842 $\pm$ 0.1116 & 0.55\\
        \hline
       \multirow{4}{*}{Free growth} & $b(z_1)\sigma_8(z_1)$ & 1.4663 $\pm$ 0.0828 & 1.4157 $\pm$ 0.0521 & \\ 
                                   & $b(z_2)\sigma_8(z_2)$ & 1.4511 $\pm$ 0.0758 & 1.5092 $\pm$ 0.0398 & \\
                                & $f(z_1)\sigma_8(z_1)$ & 0.3665 $\pm$ 0.0601 & 0.3512 $\pm$ 0.0583 & 0.4260 \\
                              & $f(z_2)\sigma_8(z_2)$ & 0.4031 $\pm$ 0.0586 & 0.4602 $\pm$ 0.0378 & 0.4367\\
        \hline
    \end{tabular}
    \caption{Constraints on parameters describing growth and clustering bias of
    galaxies with respect to the matter field in different models with and
    without including measurements from scales more than $60\mpcoh$.
    ``Standard'' model refers to the spatially-flat $\Lambda$CDM with
    $\Omega_{\rm m}$, $\sigma_8=0.8$  and general relativity.}
      \label{tab:cons}
  \end{table*}
\end{center}

We use the method outlined in Sec.~\ref{sec:meas} to constrain
parameters describing the redshift evolution of the clustering of
LRGs. We first use only scales up to $60\mpcoh$ and then use all
scales up to $200\mpcoh$. Our results are presented in
Tab.~\ref{tab:cons}.

For the wCDM + GR model the small scale data constrains the real-space
amplitude of the galaxy clustering signal in both redshift bins with
the accuracy of about 5 percent. These measurements are consistent with
previous estimates, showing that LRGs are highly biased tracers of the
underlying matter field. The power spectrum amplitudes in the two
redshift bins are close and consistent with the assumption of the
constant clustering amplitude. The constraints improve when we
extend the analysis to include data on scales 60 -- 200$\mpcoh$. This
inclusion results in slightly higher estimates of bias in higher 
redshift bin, but the two measurements are consistent at a 1$\sigma$
confidence level.

For the more general $\gamma$ parametrization that has GR as a
specific case, the $\gamma$ parameter is constrained with the precision of about
22 percent with small scale data and 19 percent when larger scales are included.
Both large and small scale measurements prefer a weaker growth than in GR but
are consistent with GR results at 1$\sigma$ confidence level. When the large
scale data is included the best-fit values for $\gamma$ are closer to the GR
values. The $\gamma$ parametrization fits give best-fit values of bias in both
redshift bins that are consistent with those measured assuming GR,
following the standard degeneracy between the bias and the RSD signal. 

Note that when adopting the $\gamma$ parametrization, we have
implicitly assumed that the growth is modified with respect to GR in a
manner that is independent of the scale. Even if the real
modifications of gravity are scale dependent, the $\gamma$
parametrization will still be able to capture deviations from GR, but
the measured $\gamma$ will be an average over the scales being
considered.

For the most general model of free growth the parameter $f\sigma_8$ can be
measured with the accuracy of about 15 percent in both redshift bins. The
inclusion of large scale data, again improves these constraints slightly.  This
shows that at larger scales SDSS DR7 clustering data is noisier and introduces
more scatter.  The best fit values of growth, when it is allowed to freely vary
are consistent with the predictions of GR. Recovered values of bias are
consistent with the ones measured in previous two models. These measurements
should be considered as giving the average growth over the scales covered.

Similar measurements for $b\sigma_8$ and $f\sigma_8$ based on fits to measured
two-dimensional $\xi^{\rm s}(r,\mu)$ have been made before from SDSS DR6
\citep{cabre09} and SDSS DR7 \citep{song10} based catalogs. Our constraints on
bias (about 5 percent in both redshift bins) are comparable to
previously derived results (about 6 percent at lower redshift and 8 percent at higher
redshift in \citet{cabre09}; about 3 percent in both redshift ranges in
\citet{song10}), while our constraints on $f\sigma_8$ (about 16 percent at lower redshift
and 15 percent at higher redshift) are stronger than results derived in previous
studies (19 percent at lower redshifts and 22 percent at higher redshifts in
\citet{cabre09}; 15 percent at lower redshfits and 12 percent at higher redshift in
\citet{song10}). Slightly improved sensitivity to $f\sigma_8$ could be due to
the fact that we are fitting to Legendre momenta function rather
than the two-dimensional correlation function itself.

The measurements of the pair of variables $b\sigma_8$, $f\sigma_8$ and
$b\sigma_8$, $\gamma$ are very weakly correlated and can be assumed to be
independent for all practical purposes.

\section{Conclusions}
\label{sec:conc}

In this paper we have considered systematic deviations from the linear
plane-parallel RSD model for the large scale clustering of galaxies. These
include systematic deviations due to wide-angle and non-linear effects, and
problems caused by inaccurate modelling of the redshift distribution. By testing
different models against the measurements from N-body simulations we checked
that, by including these effects, we can fit simulated large-scale RSD data
extremely well. We have also considered the relative importance of these
effects, showing that the wide-angle effects are small for the SDSS DR7 survey
and can be safely ignored even on scales as large as 200$\mpcoh$, but nonlinear
damping of the Baryon Acoustic Oscillation (BAO) peak \citep{meikson99} has to
be taken into account in order to properly fit the data. The non-isotropic
$\mu$-distribution effects are small for SDSS DR7 but much larger compared to
wide-angle effects and will be important for future surveys. 

Currently available measurements of galaxy clustering are low signal to noise on
very large scales and the inclusion of systematics that we discussed above
do not bias the estimates of growth and bias. For the next generation of surveys
(e.g., {\em BOSS}, {\em EUCLID}) the effects of uncertainties in radial selection,
large-scale nonlinearities and non-flat $\mu$-distribution will be comparable or
even larger than statistical errors.

We did not account for FOG effect in our fits. For the spherically averaged
correlation function FOG is expected to be small on the scales that we consider
but for the quadrupole of correlation function this effect could be comparable
to the errorbars on the scales up to 40 $\mpcoh$ (See,
Fig.~\ref{fig:xisum}).\footnote{For Q the effect is very large even on large
scales since the normilized quadrupole depends on the integral over correlation
function over all scales.} Since we use Eq.~(\ref{eq:nlbao}), which already
includes FOG effects, to model large scale nonlinearities, applying additional
FOG damping term would not be consistent. For the analysis of future high
precision measurements of clustering, proper modelling of FOG effects on small
and intermediate scales will be necessary.

In our analysis we only kept linear order terms in real-to-redshift space
mapping. Recent works have demonstrated that nonlinear contribution to
Eq.~(\ref{eq:stor}) introduce additional bias in theoretical estimates of
$P_{\theta\delta}$ and $P_{\delta\theta}$ and therefore in $\xi_\ell(r)$
\citep{taruya11,tang11,reid11}. Future measurements will also require careful
treatment of these nonlinear effects.

Different ways of extracting RSD information have been considered
before, including a fit to the two-dimensional correlation function
$\xi(\sigma,\pi)$ and the normalised quadrupole $Q$. We argue that the
best approach is to perform a joint fit to measured Legendre momenta
of the correlation function. Based on the simple linear plane-parallel
model they contain exactly the same information as $\xi(\sigma,\pi)$
and their measurement errors are more Gaussian, which makes the
interpretation of $\xi_\ell$ straightforward. Compared to only using the
normalised quadrupole $Q$, they contain significantly more information
and allow for the measurements of bias and growth independently
instead of measuring only their ratio. Also $Q$ was originally
proposed because it was believed to have certain advantages of being
independent of the shape of the power-spectrum and nonlinear effects.
We show that that does not hold on large scales: $Q$ is affected by
nonlinearities as much as the correlation itself and it is still
affected by AP effect and the dependence on the shape of the
power-spectrum and on the background cosmological parameters is not
completely removed.

We have analysed the SDSS DR7 LRG clustering in redshift-space and obtained
constraints on bias and parameters describing structure growth in two redshift
bins. We have presented what we consider to be a very robust analysis, taking
into account all of the effects that could influence the redshift-space
correlations function. The inclusion of the very large scale data does not
improve our measurements of bias and growth parameters significantly: current
measurements of the correlation function on the scales larger than $60\mpcoh$
are too noisy to be of practical interest. The next generation of ongoing and
planned surveys such as {\it BOSS}; \citealt{schlegel09}), {\it BigBOSS} and the
ESA {\em Euclid} mission \citep{laureijs09} will enable us to measure clustering
properties of galaxies on very large scales with high accuracy. For these
surveys where the measurements are more precise, the full treatment of RSD
effects will be very important. In addition, there will be significantly more
information available on the largest scales.

\section*{Acknowledgements}

We thank anonymous referee whose comments and suggestions helped us to improve
our paper.  LS is grateful for support from the European Research Council.  WJP
is grateful for support from the UK Science and Technology Facilities Council,
the Leverhulme trust and the European Research Council.  AR is grateful for the
support from a UK Science and Technology Facilities Research Council (STFC) PhD
studentship. We thank Beth Reid for useful comments on an early draft of this
paper and Marc Manera for useful discussions. LS acknowledges partial support
from Georgian National Science Foundation grant GNSF ST08/4-442 and SNFS SCOPES
grant no. 128040. We are grateful to LasDamas project for making their mock
catalogs publicly available. Numerical computations were done on the Sciama High
Performance Compute (HPC) cluster which is supported by the ICG, SEPNet and the
University of Portsmouth and COSMOS consortium supercomputer within the DiRAC
facility jointly funded by STFC, the Large Facilities Capital Fund of BIS and
the University of Portsmouth.


\begin{thebibliography}{99}

  \bibitem[\protect\citeauthoryear{Abazajian et al.}{2009}]{DR7} 
    Abazajian, K.~N., et al., 2009, ApJS, 182, 543
  \bibitem[\protect\citeauthoryear{Alcock \& Paczynski}{1979}]{alcock79}
    Alcock C., Paczynski B., 1979, Nature, 281, 358
  \bibitem[\protect\citeauthoryear{Ballinger, Peacock \& Heavens}{1996}]{ballinger96}
    Ballinger W.~E., Peacock J.~A., Heavens A.~F., 1996, MNRAS, 282, 877
  \bibitem[\protect\citeauthoryear{Barrow, Bhavsar \& Sonoda}{1984}]{barrow84}
    Barrow J.~D., Bhasvar S.~P., Sonoda D.~H., 1984, MNRAS, 210, 19
  \bibitem[\protect\citeauthoryear{Bharadwaj}{1996}]{bharadwaj96} 
    Bharadwaj S., 1996, ApJ, 472, 1
  \bibitem[\protect\citeauthoryear{Cabr\'{e} et al.}{2008}]{cabre08}
    Cabr\'{e} A., Fosalba P., Gazta\~{n}aga E., Manera M., 2008, MNRAS, 381, 1347
  \bibitem[\protect\citeauthoryear{Cabr\'{e} \& Gazta\~{n}aga}{2009}]{cabre09} 
    Cabr\'{e} A., Gazta\~{n}aga E., 2009, MNRAS, 393, 1183
  \bibitem[\protect\citeauthoryear{Carlson, White \& Padmanabhan}{2009}]{carlson09}
    Carlson J., White M., Padmanabhan N., 2009, PRD, 80, 043531
  \bibitem[\protect\citeauthoryear{Cole et al.}{1995}]{cole95}
    Cole S., Fisher K.B., Weinberg D.H., 1995, MNRAS, 275, 515 
  \bibitem[\protect\citeauthoryear{Corder \& Foreman}{2009}]{corder09}
    Corder G.~W., Foreman D.~I., 2009, Nonparametric statistics for non-statisticians: A step-by-step approach, Wiley-Blackwell, Hoboken, NJ 
  \bibitem[\protect\citeauthoryear{Crocce \& Scoccimarro}{2006}]{crocce06} 
    Crocce M., Scoccimarro R., 2006, PRD73, 063519
  \bibitem[{{Colless et al.}(2003)}]{colless03}
    Colless M., et al., 2003, preprint [astro-ph/0306581]
  \bibitem[\protect\citeauthoryear{Crocce \& Scoccimarro}{2008}]{crocce08} 
    Crocce M., Scoccimarro R., 2008, PRD77, 023533
  \bibitem[\protect\citeauthoryear{Crocce et al.}{2011}]{crocce11}
    Crocce M., Gazta\~{n}aga E., Cabr\'{e} A., Carnero A., S\'{a}nchez E., 2011,
    MNRAS, 417, 2577
  \bibitem[\protect\citeauthoryear{Dalal et al.}{2008}]{dalal08}
    Dalal N., Dor\'{e} O., Huterer D., Shirokov A., 2008, PRD, 77, 123514
  \bibitem[\protect\citeauthoryear{Desjacques \& Seljak}{2010}]{desjacquesseljak10}
    Desjacques V., Seljak U., 2010, Adv. Astron., 2010, 908640
  \bibitem[\protect\citeauthoryear{Desjacques, Seljak \& Iliev}{2010}]{desjacquesseljakiliev10}
    Desjacques V., Seljak U., Iliev I.T., 2010, MNRAS, 	396, 85
  \bibitem[\protect\citeauthoryear{Eisenstein et al.}{2001}]{eisenstein01} 
    Eisenstein D.J., et al., 2001, AJ, 122, 2267
  \bibitem[\protect\citeauthoryear{Eisenstein, Seo \& White}{2007}]{eisenstein07} 
    Eisenstein D.J., Seo H.-J., White M., 2007, ApJ, 665, 14
  \bibitem[\protect\citeauthoryear{Feldman, Kaiser \& Peacock}{1994}]{FKP}
    Feldman H.A., Kaiser N., Peacock J.A., 1994, ApJ, 426, 23
  \bibitem[\protect\citeauthoryear{Feldman, Watkins \& Hudson}{2010}]{feldman10}
    Feldman H.~A., Watkins R., Hudson M.~J., 2010, MNRAS, 407, 2328
  \bibitem[\protect\citeauthoryear{Fisher}{1995}]{Fis95}
    Fisher K.B., 1995, ApJ, 448, 494
  \bibitem[\protect\citeauthoryear{Fukugita et al.}{1996}]{F} 
    Fukugita, M., Ichikawa, T., Gunn, J.~E., Doi, M., Shimasaku, K., Schneider, D.~P., 1996, AJ, 111, 1748
  \bibitem[\protect\citeauthoryear{Gunn et al.}{1998}]{C} 
    Gunn, J.~E., et al., 1998, AJ, 116, 3040
  \bibitem[\protect\citeauthoryear{Guzzo et al.}{2008}]{guzzo08} 
    Guzzo L.  et al., 2009, Nat., 541, 451
  \bibitem[\protect\citeauthoryear{Hamilton}{1992}]{hamilton92} 
    Hamilton A.J.S., 1992, ApJ, 385, L5
  \bibitem[\protect\citeauthoryear{Hamilton}{1997}]{hamilton97} 
    Hamilton A.J.S., 1997, preprint (arXiv:astro-ph/9708102)
  \bibitem[\protect\citeauthoryear{Hawkins et al.}{2003}]{hawkins03} 
    Hawkins E., et al., 2003, MNRAS, 346, 78
  \bibitem[\protect\citeauthoryear{Jackson}{1972}]{jackson72}
    Jackson J.C., 1972, MNRAS, 156, 1
  \bibitem[\protect\citeauthoryear{Jennings, Baugh \& Pascoli}{2011}]{jennings11}
    Jennings E., Baugh C.~M., Pascoli C., 2011, ApJ, 727, L9 
  \bibitem[\protect\citeauthoryear{Jing \& B\"{o}rner}{2004}]{Jing04}
    Jing Y.P., B\"{o}rner G., 2004, ApJ, 617, 782
  \bibitem[\protect\citeauthoryear{Kaiser}{1987}]{kaiser87} 
    Kaiser N., 1987, MNRAS, 227, 1 
  \bibitem[\protect\citeauthoryear{Kashlinsky et al.}{2009}]{kashlinsky09}
    Kashlinsky A., Atrio-Barandela F., Kocevski D., Ebeling H., 2009, ApJ, 686, L49
  \bibitem[\protect\citeauthoryear{Kazin et al.}{2010}]{kazin10}
    Kazin E., et al., 2010, ApJ, 710, 1444
  \bibitem[\protect\citeauthoryear{Kazin, Sanchez \& Blanton}{2011}]{kazin11}
    Kazin E., Sanchez A.G., Blanton M.R., 2011, MNRAS, in press
  \bibitem[\protect\citeauthoryear{Landy \& Szalay}{1993}]{landy93} 
    Landy, S. D., Szalay, A. S., 1993, ApJ, 412, 64
  \bibitem[\protect\citeauthoryear{Laureijs et al.}{2009}]{laureijs09} 
    Laureijs R. et al., 2009, {\em Euclid} Assessment Study Report for the ESA Cosmic Visions, (arXiv:0912.0914 [astro-ph.CO]) 
  \bibitem[\protect\citeauthoryear{Lewis, Challinor \& Lasenby}{2000}]{camb}
    Lewis A., Challinor A., Lasenby A., 2000, ApJ, 538, 473
  \bibitem[\protect\citeauthoryear{Li et al.}{2007}]{Li07}
    Li C., Jing Y.P., Kauffmann G., B\"{o}rner G., Xi K., Wang L., 2007, MNRAS, 376, 984
  \bibitem[\protect\citeauthoryear{Linder}{2005}]{linder05}
    Linder E., 2005, PRD, 72, 043529
  \bibitem[\protect\citeauthoryear{Lucey}{1979}]{lucey79}
    Lucey J.~R., 1979, PhD thesis, Univ. Sussex
  \bibitem[\protect\citeauthoryear{Macaulay et al.}{2011}]{macaulay11}
    Macaulay E., Feldman H., Ferreira P.~G., Hudson M.~J., Watkings R., 2011, MNRAS, 414, 621
  \bibitem[\protect\citeauthoryear{Matsubara}{2004}]{matsubara04} 
    Matsubara T., 2004, ApJ, 615, 573
  \bibitem[\protect\citeauthoryear{Matsubara}{2008a}]{matsubara08a} 
    Matsubara T., 2008a, PRD, 77, 063530
  \bibitem[\protect\citeauthoryear{Matsubara}{2008b}]{matsubara08b} 
    Matsubara T., 2008b, PRD, 78, 083519
  \bibitem[\protect\citeauthoryear{Meikson, White \& Peacock}{1999}]{meikson99}
    Meikson A., White M., Peacock J.~A., 1999, MNRAS, 304, 851
  \bibitem[\protect\citeauthoryear{Mo, Jing \& B\"{o}rner}{1992}]{mo92}
    Mo H.~J., Jing Y.~P., B\"{o}rner G., 1992, ApJ, 392, 452
  \bibitem[\protect\citeauthoryear{Norberg et al.}{2009}]{norberg09}
    Norberg P., Baugh C.M., Gazta\~naga E., Croton D.J., 2009, MNRAS, 396, 19
  \bibitem[\protect\citeauthoryear{Nusser \& Davis}{2011}]{nusser10}
    Nusser A., Davis M., 2011, ApJ, 736, 93
  \bibitem[\protect\citeauthoryear{Okamura \& Jing}{2011}]{okamura11}
    Okamura T., Jing Y.~P., 2011, ApJ, 726, 11
  \bibitem[\protect\citeauthoryear{Okumura et al.}{2008}]{okumura08}
    Okumura T., Matsubara T., Eisenstein D.J., Kayo I., Hikage C., Szalay A.S., Schneider D.P., 2008, ApJ, 676, 889
  \bibitem[\protect\citeauthoryear{Padmanabhan et al.}{2005}]{padmanabhan05}
    Padmanabhan N., et al., 2005, PRD, 72, 043525
  \bibitem[\protect\citeauthoryear{Papai \& Szapudi}{2008}]{papai08} 
    Papai P., Szapudi I., 2008, MNRAS, 389, 292
  \bibitem[\protect\citeauthoryear{Peacock \& Dodds}{1996}]{PeaDod96}
    Peacock J.A., Dodds S.J., 1996, MNRAS, 280, L19
  \bibitem[\protect\citeauthoryear{Percival et al.}{2004}]{percival04} 
    Percival W. J. et al., 2004, MNRAS, 353, 1201
  \bibitem[\protect\citeauthoryear{Percival et al.}{2010}]{percival10} 
    Percival W. J. et al., 2010, MNRAS, 401, 2148
  \bibitem[\protect\citeauthoryear{Press et al.}{1992}]{press92} 
    Press W.H., Teukolsky S,A., Vetterling W.T., Flannery B.P., 1992, 
    Numerical recipes in C. The art of scientific computing, Second edition, 
    Cambridge: University Press. 
  \bibitem[\protect\citeauthoryear{Raccanelli, Samushia \& Percival}{2010}]{raccanelli10} 
    Raccanelli A., Samushia L., Percival W. J., 2010, MNRAS, 409, 1525
  \bibitem[\protect\citeauthoryear{Reid et al.}{2009}]{reid09} 
    Reid B. A., et al., 2010, MNRAS, 404, 60
  \bibitem[\protect\citeauthoryear{Reid \& White}{2011}]{reid11}
    Reid B. A., White M., 2011, MNRAS, 417, 1913
  \bibitem[\protect\citeauthoryear{Ross et al.}{2011}]{ross11}
    Ross A. J., Percival W. J., Crocce M., Cabr\'{e} A., Gazga\~{n}aga E., 2011,
    MNRAS, 415, 2193
  \bibitem[\protect\citeauthoryear{Samushia et al.}{2010}]{samushia10} 
    Samushia L., et al. 2010, MNRAS, 410, 1993
  \bibitem[\protect\citeauthoryear{Sanchez et al.}{2009}]{sanchez09}
    Sanchez A., Crocce M., Cabr\'{e} A., Baugh C.M., Gazta\~{n}aga E., 2009, MNRAS, 400, 1643
  \bibitem[\protect\citeauthoryear{Schlegel, White \& Eisenstein}{2009}]{schlegel09} 
    Schlegel D., White M., Eisenstein D.J., 2009, The astronomy and astrophysics decadal survey, science white paper No. 314, (arXiv:0902.4680 [astro-ph.CO])
  \bibitem[\protect\citeauthoryear{Schlegel et al.}{2009}]{schlegeletal09}
    Schlegel et al., 2009, preprint (arXiv:0904.0468 [astro-ph.CO])
  \bibitem[\protect\citeauthoryear{Scoccimarro}{2004}]{scoccimarro04} 
    Scoccimarro R., 2004, PRD, 70, 083007
  \bibitem[\protect\citeauthoryear{Seljak, Hamaus \& Desjacques}{2009}]{seljak09} 
    Seljak U., Hamaus N., Desjacques V., 2009, PRL 103, 091303
  \bibitem[\protect\citeauthoryear{Simpson \& Peacock}{2010}]{simpson10} 
    Simpson F., Peacock J., 2010, PRD, 81, 043512
  \bibitem[\protect\citeauthoryear{Smith et al.}{2003}]{smith03}
    Smith R.~E., Peacock J.~A., Jenkins A., White S.~D.~M., et al., 2003, MNRAS, 341, 1311
  \bibitem[\protect\citeauthoryear{Song et al.}{2011}]{song10} 
    Song Y.-S., Sabiu C., Kayo I., Nichol B., 2011, JCAP, 05, 020
  \bibitem[\protect\citeauthoryear{Szalay, Matsubara \& Landy}{1998}]{szalay98} 
    Szalay A.  S., Matsubara T., Landay S. D., 1998, ApJ, 498, L1
  \bibitem[\protect\citeauthoryear{Szapudi}{2004}]{szapudi04} 
    Szapudi I., PRD, 70, 083536
  \bibitem[\protect\citeauthoryear{Takada}{2006}]{takada06}
    Takada M., PRD, 2006, 74, 043505
  \bibitem[\protect\citeauthoryear{Tang}{2011}]{tang11}
    Tang J., Kayo I., Takada M., 2001, MNRAS, 416, 229
  \bibitem[\protect\citeauthoryear{Taruya, Nishimichi \& Saito}{2010}]{taruya10} 
    Taruya A., Nishimichi T., Saito S., 2010, PRD, 82, 063522
  \bibitem[\protect\citeauthoryear{Taruya, Saito \& Nishimichi}{2011}]{taruya11}
    Taruya A., Saito S., Nishimichi T., 2011, PRD, 83, 103527
  \bibitem[\protect\citeauthoryear{Tegmark et al.}{2006}]{tegmarketal06} 
    Tegmark M. et al., 2006, PRD, 74, 123507
  \bibitem[\protect\citeauthoryear{Thomas, Abdalla \& Lahav}{2011}]{thomas10}
    Thomas S.~A., Abdalla F.~B., Lahav O., 2011, PRL, 106, 241301
  \bibitem[\protect\citeauthoryear{Tocchini-Valentini et al.}{2011}]{tocchini11}
    Tocchini-Valentini D., Barnard M., Bennett C.~L., Szalay A.~S., 2011, preprint (arXiv:1101.2608 [astro-ph.CO])
  \bibitem[\protect\citeauthoryear{Watkins, Feldman \& Hudson}{2009}]{watkins09}
    Watkins R., Feldman H.~A., Hudson M.~J., 2009, MNRAS, 392, 743
  \bibitem[{{York et al.}(2000)}]{york00}
    York D.G., et al., 2000, AJ, 120, 1579
  \bibitem[\protect\citeauthoryear{Zaroubi \& Hoffman}{1993}]{zaroubi93} 
    Zaroubi S., Hoffman Y., 1993, preprint (arXiv:astro-ph/9311013)
  \bibitem[\protect\citeauthoryear{Zehavi et al.}{2005}]{zehavi05} 
    Zehavi I. et al., 2005, ApJ, 621, 22
  
\end{thebibliography}
\end{document}